%% file: HRSE_paper.tex
\documentclass[iop, twocolumn, tighten]{emulateapj}
\usepackage{pifont}
\usepackage{amsmath}
\usepackage{color}
\usepackage{ulem}

\definecolor{green}{rgb}{0.0,0.8,0.0}
\definecolor{red}{rgb}{0.8,0.0,0.0}

 \include{rule}

\slugcomment{}
\shortauthors{}
\shorttitle{}

\begin{document}
%%%%%%%%%%%%%%%%%%%%%%%%%%%%%%%%%%%%%%%%%%%%%%%%%%%%%%%%%%%
%\setlength{\baselineskip}{36pt}
%%%%%%%%%%%
\title{Theoretical Emission Spectra of Atmospheres of Hot Rocky Super-Earths}
%%%%%%%%%%%%%%%%%%%%%%%%%%%%%%%%%%%%%%%%%%%%%%%%%%%%%%%%%%%%%%%%%%%%%%

%%%%%%%%%%%%%%%%%%%%%%%%%%%%%%%%%%%%%%%%%%%%%%%%%%%%%%%%%%%%%%%%%%%%%%
\author{Yuichi Ito\altaffilmark{1}, Masahiro Ikoma\altaffilmark{1}, Hajime Kawahara\altaffilmark{1}, Hiroko Nagahara\altaffilmark{1}, Yui Kawashima\altaffilmark{1}, and Taishi Nakamoto\altaffilmark{2}} 
\altaffiltext{1}{Department of Earth and Planetary Science, The University of Tokyo, 7-3-1 Hongo, Bunkyo-ku, Tokyo 113-0033, Japan.}
\altaffiltext{2}{Department of Earth and Planetary Sciences, Tokyo Institute of Technology, 2-12-1 Ookayama, Meguro-ku, Toyo 152-8551, Japan.}
%\email{mail@eps.s.u-tokyo.ac.jp}
%%%%%%%%%%%%%%%%%%%%%%%%%%%%%%%%%%%%%%%%%%%%%%%%%%%%%%%%%%%%%%%%%%%%%%
\begin{abstract}
Motivated by recent detection of transiting high-density super-Earths, we explore the detectability of hot rocky super-Earths orbiting very close to their host stars. 
In the environment hot enough for their rocky surfaces to be molten, they would have the atmosphere composed of gas species from the magma oceans. 
In this study, we investigate the radiative properties of the atmosphere that is in the gas/melt equilibrium with the underlying magma ocean. 
Our equilibrium calculations yield Na, K, {Fe,} Si, SiO, O, and O$_2$ as the major atmospheric species. 
We compile the radiative-absorption line data of those species available in literature, and calculate their absorption opacities in the wavelength region  of 0.1--100~$\mathrm{\mu m}$. 
Using them, we integrate the thermal structure of the atmosphere. 
Then, we find that thermal inversion occurs in the atmosphere because of the UV absorption by SiO. 
In addition, we calculate the ratio of the planetary to stellar emission fluxes during secondary eclipse, and {find prominent emission features induced by SiO at 4~$\mathrm{\mu m}$ detectable by Spitzer, and those at 10 and 100~$\mathrm{\mu m}$ detectable by near-future space telescopes.}

\end{abstract}


%%%%%%%%%%%%%%%%%%%%%%%%%%%%%%%%%%%%%%%%%%%%%%%%%%%%%%%%%%%%%%%%%%%%%%
\section{INTRODUCTION}

We have recently entered a new era of characterizing the atmospheres of low-mass exoplanets with masses of $\lesssim$~30~$M_\oplus$. Transmission and/or emission spectra have been obtained for several low-mass exoplanets in transit \citep[e.g.,][]{Deming+07,Bean+10}. 
{Most of them are  
less dense than they would be if they were composed of silicate. }  
Characterization of the relatively thick atmospheres of 
{such} exoplanets has been paid attention to as a first stepping stone towards an observational science as to habitable planets.

A next stepping stone would be detection and characterization of thin atmospheres {of close-in low-mass exoplanets, {the observed mass-radius relationships for which are consistent with theoretical ones for rocky planets \citep[e.g.,][]{Fortney+2007, Valencia+07, Grasset+09}.
In particular, of special interest in this study are rocky super-Earths with masses of $\leq$ 10~$\Mearth$ and radii of $\leq$ 2~$\Rearth$ that are orbiting so close to their host stars that their surface temperatures are over the vaporization temperature of rock, because of intense stellar irradiation. }
Such exoplanets are}  referred to as hot rocky super-Earths (HRSEs, hereafter) {in this study}.
{Examples} of the transiting HRSEs detected so far are CoRoT-7~b \citep{Leger+09}, Kepler-10~b \citep{Batalha+11}, Kepler-78~b \citep{Howard+2013}, and {55 Cnc e \citep{Winn+11}}.
They {are likely to} have atmospheres with chemical species of rocky origin.

Based on their chemical equilibrium calculations,  
\citet{Schaefer+09} presented the possible compositions of the ``silicate atmosphere" that is in gas-melt equilibrium with the molten rocky surface with no highly volatile elements such as H, C, N, S, and Cl (i.e., volatile-free magma ocean). 
They considered temperatures, $T$, of 1500-3000~K and gravity of 36.2~$\rm m \, s^{-2}$, which corresponded to the planetary properties for CoRoT-7~b, and assumed several compositions for the volatile-free magma including the Earth's continental and oceanic crusts, the bulk silicate Earth (BSE), and the bulk silicate Moon. 
Their calculations demonstrated that the atmosphere contained Na, O, O$_2$, and SiO as the major constituents. 
Also, they investigated changes in atmospheric composition due to partial loss of the atmosphere, and found that Na was selectively lost from the atmosphere, while Mg became more abundant with fractional loss. 

\citet{Miguel+11} made a similar but more extensive investigation of the composition of atmospheres of volatile-free HRSEs. 
They considered wider ranges of temperature (1000 to 3500~K), planetary mass (1 to 10~$M_\oplus$), and planetary radius (1 to 2.5~$R_\oplus$). 
Also they considered komatiite in addition to BSE for the magma material. 
The resultant {atmospheric} compositions are basically similar to those from \citet{Schaefer+09}: 
The major constituents are Na, O, O$_2$, and SiO, although Fe and Mg are more abundant than SiO at $T \lesssim$~2000~K. 
They defined five types of HRSE atmosphere, depending on the abundance of the six species, and then classified \textit{Kepler} planet candidates with radii of $\lesssim$~2.5~$R_\oplus$, according to type. 

Since the presence of such atmospheres is specific to rocky planets, 
observational detection of those atmospheres could {add a piece of convincing} evidence for rocky planets. 
Furthermore, identifying the atmospheric constituents could give constraints on the bulk composition and formation process of HRSEs. 
However, the properties and detectability of {the atmospheres of the volatile-free HRSEs} have not been examined yet. 
Thus, the aim of this study is to identify the radiative properties of the atmospheres on top of the global magma oceans of {volatile-free} HRSEs, and then assess their detectability. 
(In this study, we call such an atmosphere a mineral atmosphere.)

To this end, we model the thermal structure of the mineral atmospheres of HRSEs. 
The composition of the atmosphere is obtained numerically in section~\ref{sec:chemistry}. 
Also, we compile the absorption line data for the major gas species, which are published in the literature or online, and then calculate the absorption opacities in section~\ref{sec:opacity}. 
Using the chemical compositions and absorption opacities, we integrate the thermal structure of the irradiated atmosphere in radiative equilibrium in section~\ref{sec:structure}. 
Then, we evaluate the detectability of the chemical species in the mineral atmosphere through the emission spectra obtained during secondary eclipse in section~\ref{sec:detectability}. 
Finally, we discuss several effects to be examined in section~\ref{sec:discussion} and then conclude this study in section~\ref{sec:concl}.

%%%
%%% SECTION 2
%%%
\section{CHEMICAL COMPOSITION} \label{sec:chemistry}
\subsection{Element abundance and gas pressure} \label{subsec:element}
\begin{figure}[htbp]
   \begin{center}
  \includegraphics[width=80mm]{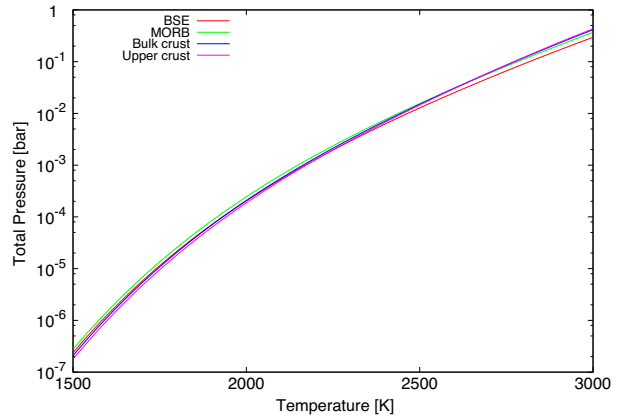}
   \end{center}
   \caption{
   Total pressures of gas in gas/melt equilibrium with magma of assumed composition are shown as functions of temperature. 
  The assumed bulk compositions are the same as those of  
   the bulk silicate Earth (BSE) (red), 
   the Earth's mid-ocean ridge basalts (MORB) (green), 
   the Earth's bulk crust (blue), and 
   the Earth's upper crust (magenta) (see Table~\ref{tbl:melt} for the details). 
   }
 \label{fig:hsep}
 \end{figure}

\begin{figure*}[htbp]
 \begin{minipage}{0.5\hsize}
  \begin{center}
   (a) BSE \\
   \includegraphics[width=0.75\linewidth]{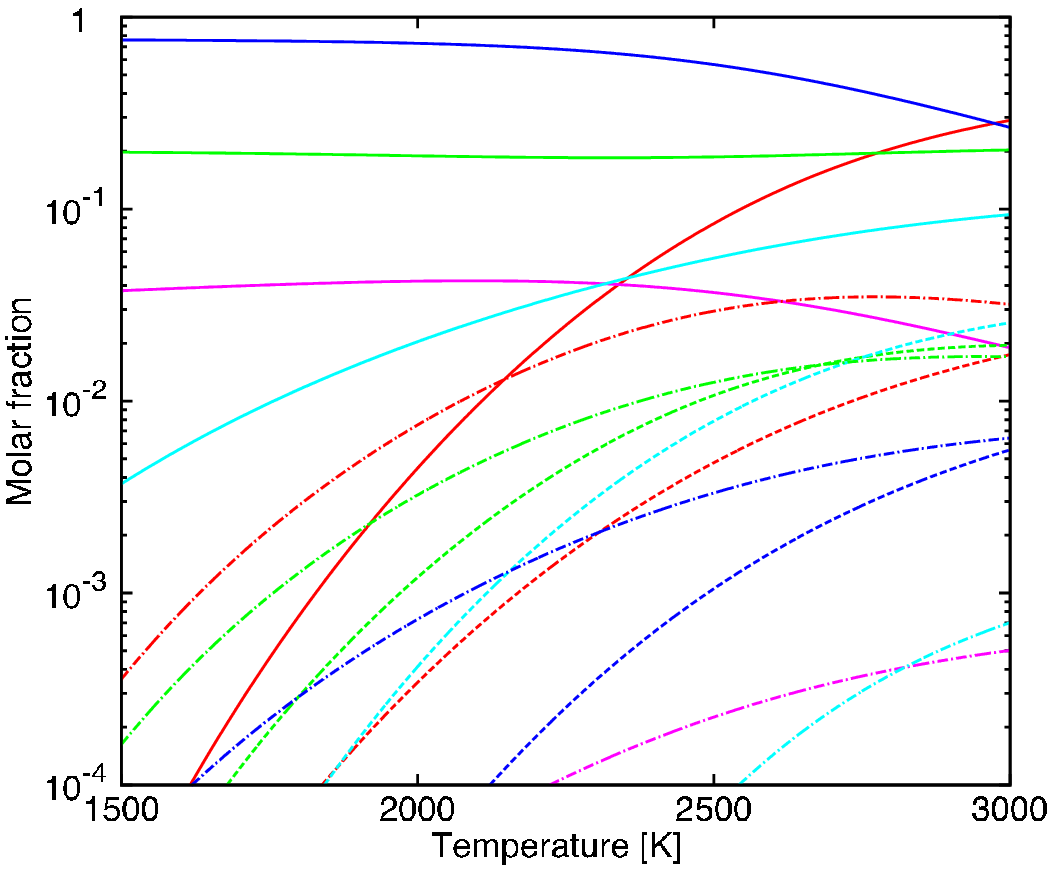}
  \end{center}
  \label{fig:bse}
 \end{minipage}
 \begin{minipage}{0.5\hsize}
  \begin{center}
   (b) MORB \\
   \includegraphics[width=0.75\linewidth]{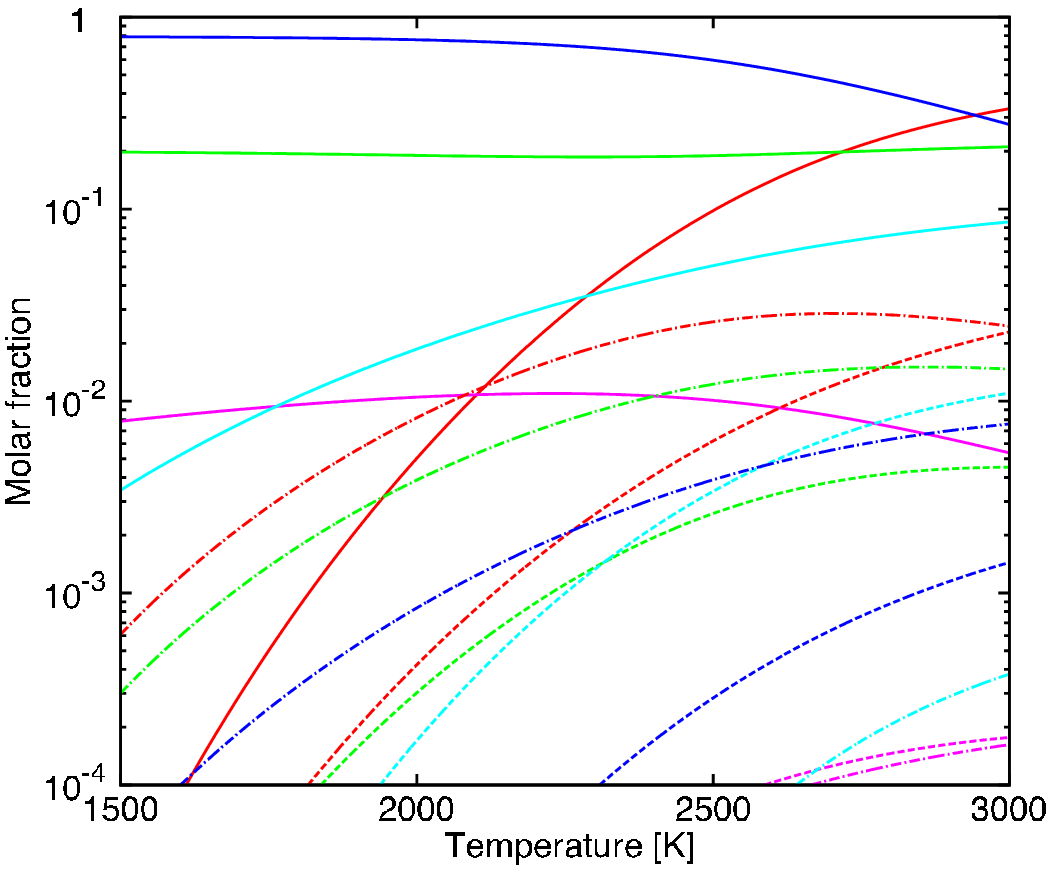}
  \end{center}
  \label{fig:morb}
 \end{minipage} 
  \begin{minipage}{0.5\hsize}
  \begin{center}
    (c) Bulk crust \\
   \includegraphics[width=0.75\linewidth]{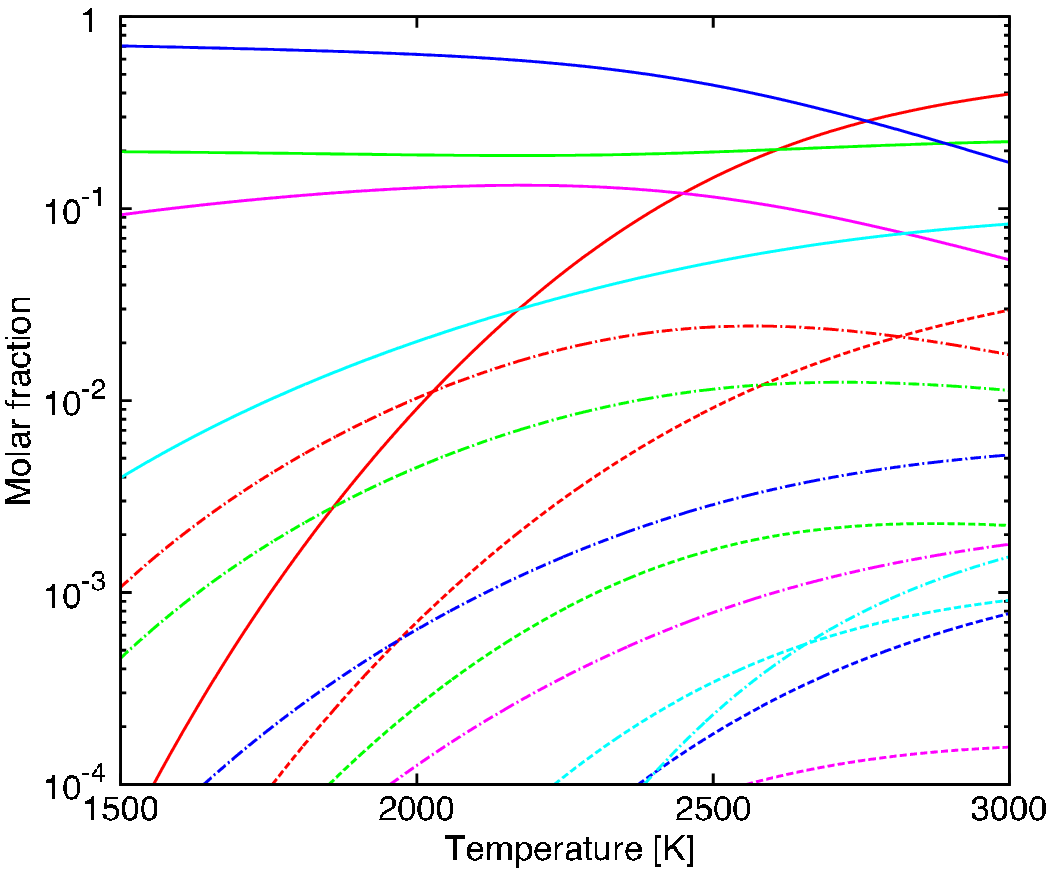}
  \end{center}
  \label{fig:bcrust}
 \end{minipage}
 \begin{minipage}{0.5\hsize}
  \begin{center}
   (d) Upper crust \\
   \includegraphics[width=0.88\linewidth]{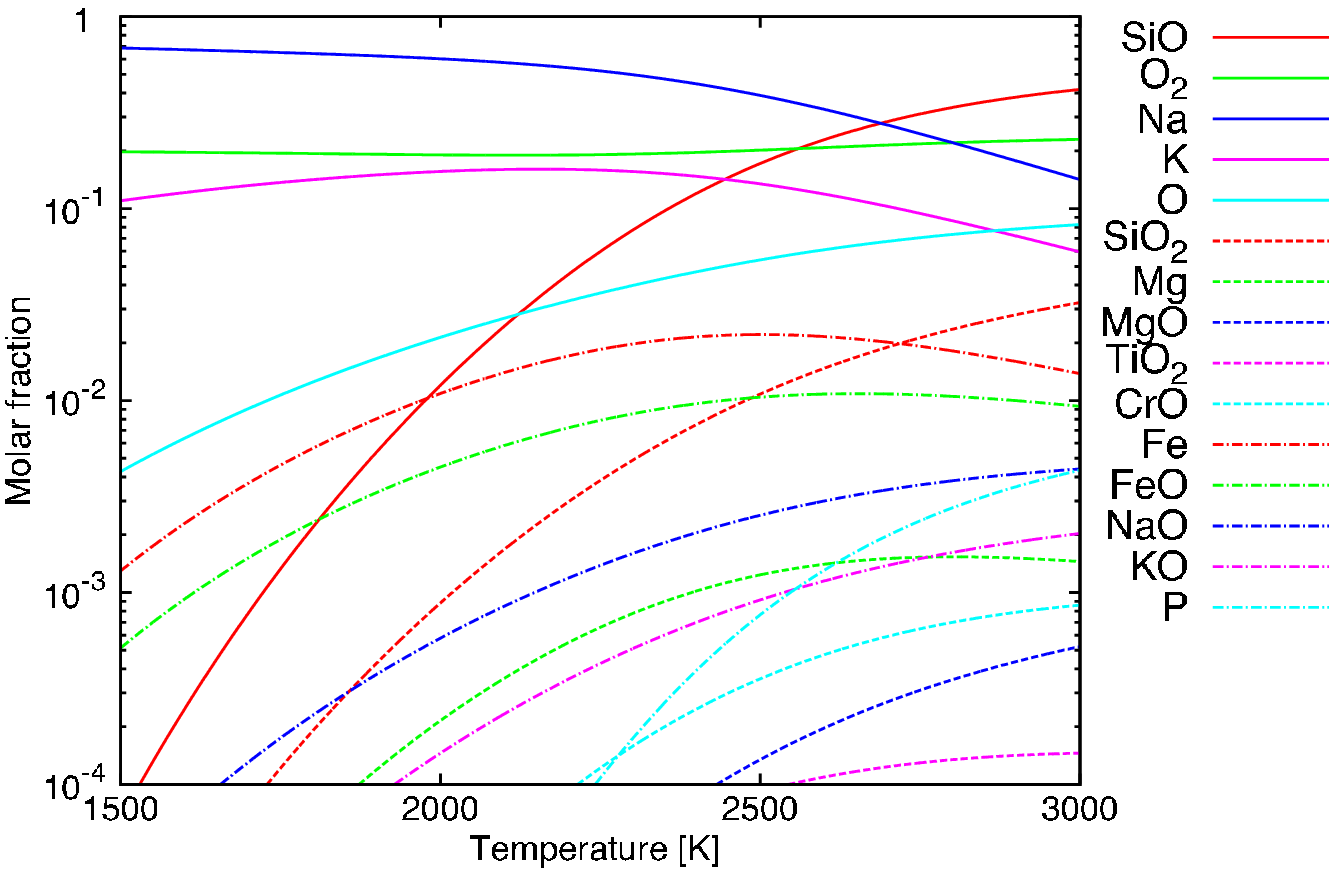}
  \end{center}
  \label{fig:ucrust}
 \end{minipage}
  \caption{
    Composition of gas in chemical equilibrium with molten silicate. 
  Molar fractions of the gas species are shown as functions of temperature. 
  The assumed bulk compositions are the same as those of  
  (a) the bulk silicate Earth (BSE), 
  (b) the Earth's mid-ocean ridge basalts (MORB), 
  (c) the Earth's bulk crust, and 
  (d) the Earth's upper crust (see Table~\ref{tbl:melt} for the details). 
}
 \label{fig:COMP_X}
\end{figure*}

To calculate the element abundance and pressure of gas in gas/melt chemical equilibrium with silicate melt for a given temperature, we use the chemical equilibrium code MELTS \citep{Ghiorso+95,Asimow+98}, which performs the free-energy minimization calculations. 
Figures~\ref{fig:hsep} and \ref{fig:COMP_X} show the calculated total pressures and molar fractions of gas species, respectively,
 over a temperature range between 1500 and 3000~K for four magma compositions such as 
 (a) BSE, (b) the Earth's mid-ocean ridge basalts (MORB), (c) the Earth's bulk crust, and (d) the Earth's upper crust. 
The detail of the magma compositions is shown in Table~\ref{tbl:melt}. 
{
In this study, we consider the above four compositions of magma, following \citet{Schaefer+09}, although the magma composition  
of exoplanets is uncertain. 
The actual magma composition depends not only on the element abundances but on the thermal history of the planet; exploring the details is beyond the scope of this study.
{Note that the choice among the four magma compositions has little impact on the emission spectra in the infra-red wavelength region.
Detectable difference may arise in the visible emission spectra, which is discussed in section~\ref{sec:diss_k}}
}

The overall trend shown in Figs.~\ref{fig:hsep} and \ref{fig:COMP_X} is similar with that found by \citet{Schaefer+09} and \citet{Miguel+11}, who used the MAGMA chemical equilibrium code \citep{Fegley+87}, instead of MELTS. 
For example, Na is the most abundant species in most of the temperature range, while SiO increases with temperature and becomes the most abundant one for $T \gtrsim 2800$~K for BSE and MORB, $T \gtrsim 2600$~K for the bulk crust, and $T \gtrsim 2500$~K for the upper crust. 
Also, the total vapor pressure is as small as $\sim$~$10^{-7}$~bar at $T$ = 1500~K and $\sim$~0.1~bar at $T$ = 3000~K, for example
 (see Fig.~\ref{fig:hsep}). 

{
A noticeable difference is seen in the abundance of potassium gas between our calculations and those by \citet{Schaefer+09} and \citet{Miguel+11}. 
Our calculations yield much higher partial pressure of potassium: 
The ratio of the potassium to sodium partial pressures that we have obtained is larger by a factor of more than 100, relative to that calculated by \citet{Schaefer+09}. 
This difference is due mainly to that in the adopted thermodynamic model for silicate melt. 
The melt model of MAGMA used in \citet{Schaefer+09} is ideal mixing of many simple and complex fictitious oxide species, which is calibrated for silicate melts with high-K$_2$O contents, whereas MELTS used in this study adopts a symmetric regular solution model with minimum melt components, which is calibrated by experimental results on natural systems with low-K$_2$O abundances. 
MAGMA would appropriately estimate potassium vapor pressure for high-K$_2$O melt as shown in Fig.~5 of \citet{Schaefer+04}, but is not good for low-K$_2$O melt.  
By contrast, MELTS gives appropriate results for low-K$_2$O melt, which is of interest in this study (see Table~\ref{tbl:melt}).  
The impact on the emission spectrum from the mineral atmosphere is discussed in section~\ref{sec:diss_k}.
}

\begin{table}[h]
\caption{Bulk compositions considered in this study}
    \label{tbl:melt}
  \begin{tabular}{l r r r r } \hline \hline
    Oxide(wt\%) & BSE$^a$ & MORB$^a$ & Bulk crust$^b$ & Upper crust$^b$ \\ \hline 
    SiO$_2$ & 45.1 & 49.6 & 60.6 & 66.6 \\
    MgO & 37.9 & 9.75 & 4.70 & 2.50 \\
    FeO & 8.06 & 8.06 & 6.71 & 5.04 \\ 
  Al$_2$O$_3$ & 4.46 & 16.8 & 15.9 & 15.4 \\
    CaO & 3.55 & 12.5 & 6.41 & 3.59 \\
    Na$_2$O & 0.36 & 2.18 & 3.07 & 3.27 \\
    Cr$_2$O$_3$ & 0.38 & 0.07 & 0.0004 & 0.0003 \\
    TiO$_2$ & 0.20 & 0.9 & 0.72 & 0.64 \\ 
    K$_2$O & 0.03 & 0.07 & 1.81 & 2.80 \\
    P$_2$O$_5$ & 0.02 & 0.10 & 0.13 & 0.15\\ \hline
  \end{tabular}  
  
  Notes.
  \footnotemark{\citet{McDonough+95}}, \footnotemark{\citet{Rudnick+03}}
\end{table}

\subsection{Gas phase equilibrium} \label{subset:gas}
To calculate the equilibrium abundances of gas species in the mineral atmosphere, we use the NASA CEA code \citep{Gordon+96}, which performs the free-energy minimization calculations. 
In practice, we calculate the abundances for some selected temperatures from 2000~K to 4500~K and pressures from 10$^{-8}$ bar to 10$^{-1}$ bar, which are of interest in this study, considering the gas phase equilibrium of a system composed of only the major elements such as O, Na, K, Si and {Fe}. 
We assume that the element abundances are vertically constant{; their} values are calculated in section~\ref{subsec:element}.
    
In our atmosphere model, the molar fractions of Na, K {and Fe} gas vary little, compared to those of O, O$_2$, Si and SiO.
O$_2$ and SiO tend to decrease and O and Si increase with altitude in low pressure regions in the atmosphere (see section \ref{sec:structure}), because of thermal dissociations of O$_2$ and SiO.
  
Although we ignore photo-chemical reactions in this study, the photo-dissociation of SiO may be important for the detectability of the atmosphere. We discuss the possibility of the photo-dissociation of SiO and the impact on the atmospheric structure and detectability in section~\ref{subsec:photod}.

%%%
%%% SECTION 3
%%%  
\section{ABSORPTION OPACITY} \label{sec:opacity}

\begin{figure*}[htbp]
 \begin{minipage}{0.50\hsize}
    \begin{center}
  \includegraphics[width=80mm]{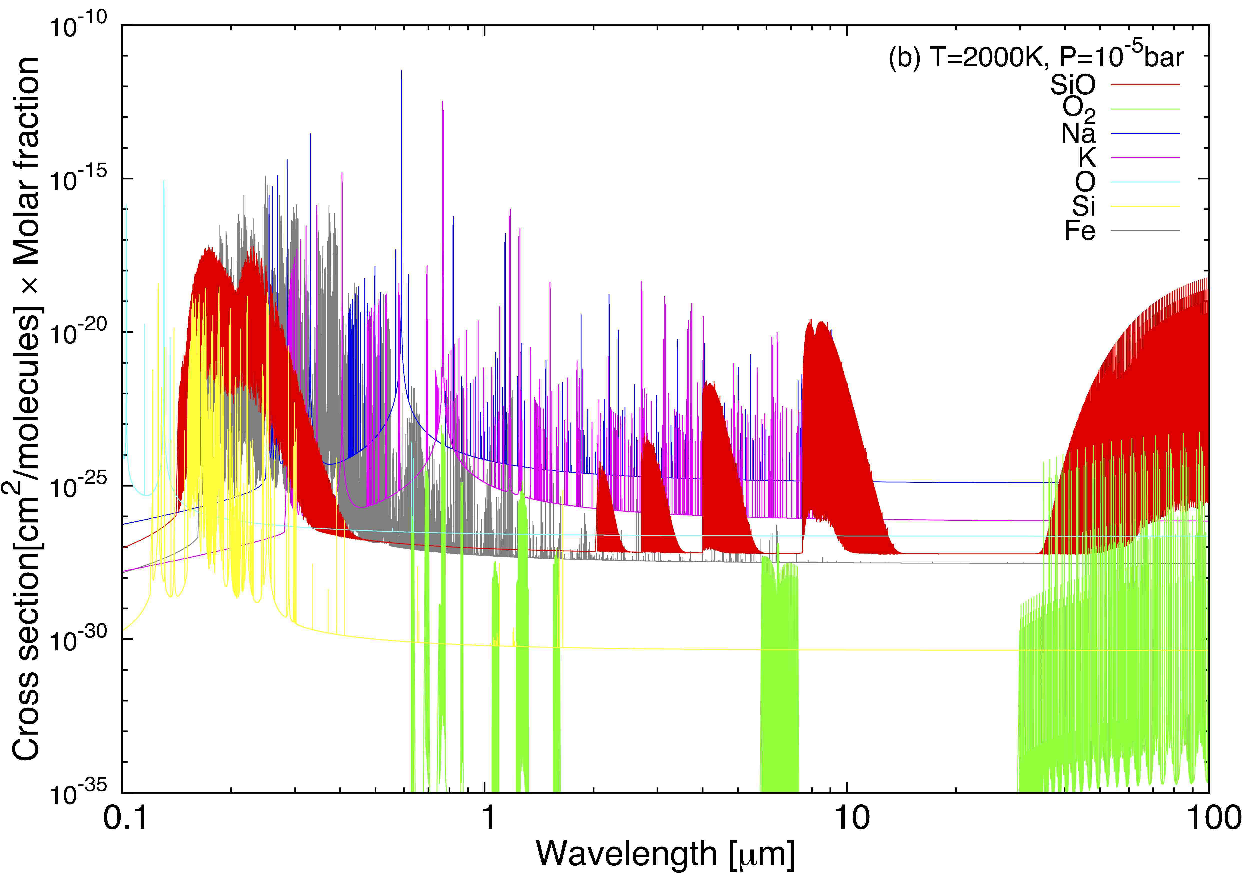}
    \end{center}
%    \caption{
%    Absorption cross sections of the five major gas species, SiO (red), 
%    O$_2$ (green), Na (blue), K (magenta), and O (cyan), as functions of wavelength 
%    for temperature of 2000~K and pressure of $1 \times 10^{-5}$~bar. 
%    Each cross section is multiplied by each molar fraction. 
%    The magma composition is assumed to be BSE (see Table~\ref{tbl:melt}). 
%    The calculated molar fractions of Na, K, SiO, O, and O$_2$ are 
%    $0.721$, $0.0407$, $0.00731$, $0.0233$, and $0.189$, respectively.
%    }
% \label{fig:2000c}
 \end{minipage}
     \begin{minipage}{0.50\hsize}
   \begin{center}
  \includegraphics[width=80mm]{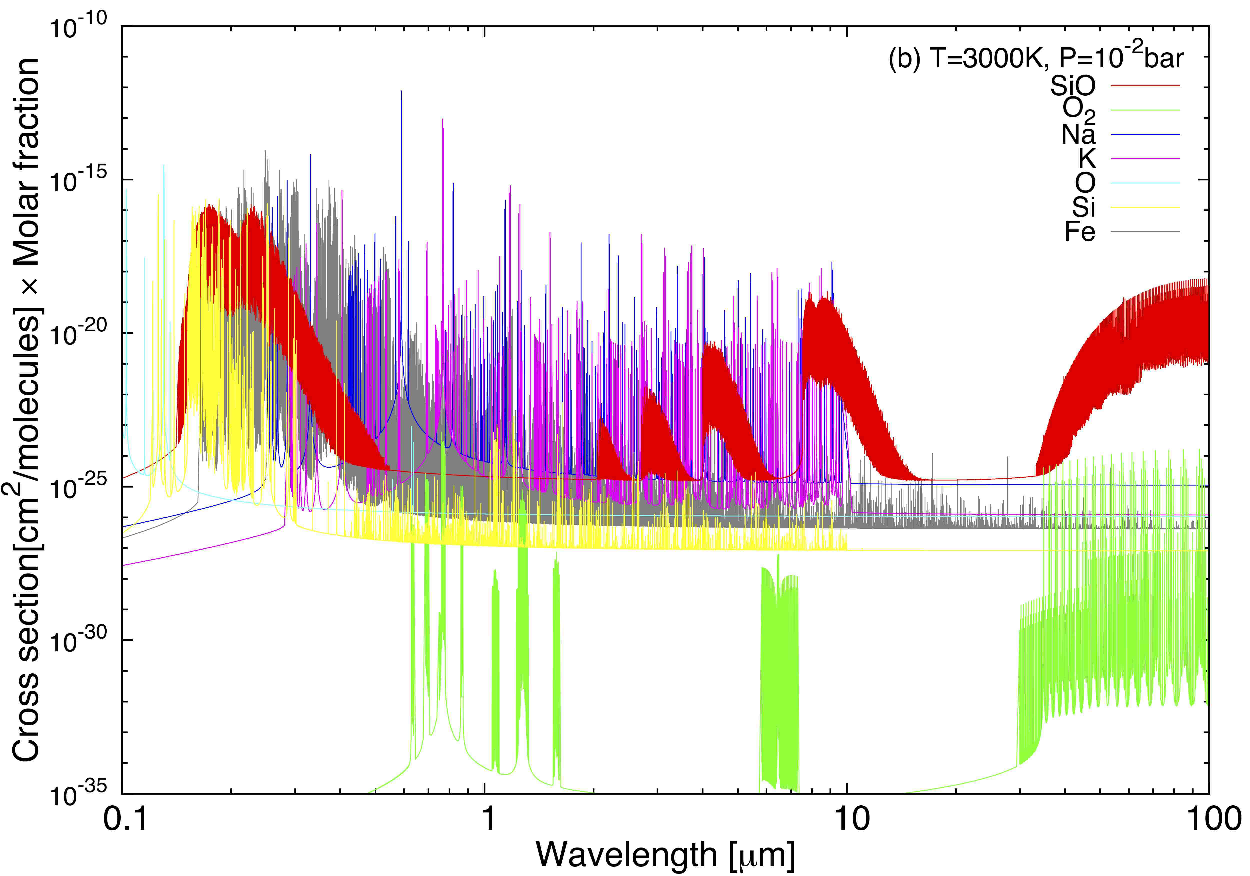}
   \end{center}
%   \caption{
%   Absorption cross sections of the five major gas species, SiO (red), 
%   O$_2$ (green), Na (blue), K (magenta), and O (cyan), as functions of wavelength 
%   for temperature of 3000~K and pressure of $1 \times 10^{-2}$~bar. 
%   Each cross section is multiplied by each molar fraction. 
%   The magma composition is assumed to be BSE (see Table~\ref{tbl:melt}). 
%   The calculated molar fractions of Na, K, SiO, O, and O$_2$ are 
%   $0.204$, $0.0142$, $0.343$, $0.0991$, and $0.211$, respectively.
%   }
% \label{fig:3000c}
 \end{minipage}
 \caption{Absorption cross sections of the seven major gas species, SiO (red), 
   O$_2$ (green), Na (blue), K (magenta), Fe (grey), O (cyan), and Si(yellow) as functions of wavelength.
       Each cross section is multiplied by each molar fraction. 
    The magma composition is assumed to be BSE (see Table~\ref{tbl:melt}).
    (a): The temperature is 2000~K and the pressure is $1 \times 10^{-5}$~bar.
    The calculated molar fractions of Na, K, Fe, Si, SiO, O, and O$_2$ are 
    $0.721$, $0.0407$, $0.001$, $1 \times 10^{-6}$, $0.00731$, $0.0233$, and $0.189$, respectively.
    (b): The temperature is 3000~K and the pressure is $1 \times 10^{-2}$~bar.
    The calculated molar fractions of Na, K, Fe, Si, SiO, O, and O$_2$ are 
   $0.204$, $0.0142$, $0.01$, $1 \times 10^{-3}$, $0.343$, $0.0991$, and $0.211$, respectively. 
    }
 \label{fig:cross}
 \end{figure*}

Among the chemical species shown in Figure~\ref{fig:COMP_X}, we consider line absorption by {seven} major species that include Na, K, {Fe,} O, O$_2$, Si and SiO. 
We neglect Rayleigh scattering, because the Rayleigh scattering opacities are mostly smaller than the absorption opacity of SiO at wavelengths of interest in this study, as discussed in section~\ref{Scattering}. 
Note that while TiO is known to be a strong absorber for visible ray \citep{Allard+00,Plez98}, its absorption opacity is always smaller than Na absorption opacity in the visible and IR wavelength regions, because the molar fraction of TiO is negligibly small (e.g., $\sim 10^{-12}$ at $T$ = 1500~K and $\sim 10^{-6}$ at $T$ = 3000~K), compared to that of Na ($\gtrsim$ 0.1) in the atmosphere considered here.

Absorption opacities are calculated in the following way, except for O$_2$, for which we apply
vibrational-rotational transitions in
 the HITRAN database \citep{Rothman+13}. 
In local thermodynamic equilibrium, the absorption cross section of species A for an energy transition from the $i$th level to the $j$th level, $\sigma_{\nu}^\mathrm{A}(i,j)$, is given as \citep{Piskunov+01}
\begin{align}
\sigma^\mathrm{A}_{\nu}(i,j)= & g_i f_{ij} 
			\frac{\pi e^2}{m_e c}
			\frac{\exp({-E_i/kT})}{Q(T)} \notag \\
			&\times \left\{ 1-\exp\left(\frac{E_j-E_i}{kT}\right) \right\} 
			H(\nu,a), 
			\label{eq:sg}
			\end{align}
where $\nu$ is the frequency, $g_i$ is the statistical weight of the $i$th energy level,
 $f_{ij}$ is the oscillator strength for a transition from the $i$th (lower) to $j$th (higher) level,
$E_i$ and $E_j$ are excitation energies of the $i$th and $j$th energy levels, respectively,
 and $Q(T)$ is the partition function of the species at temperature $T$.
$e$ is the elementary charge (= $4.803 \times10^{-10}$~esu), $m_e$ is the electron mass (= $9.1094 \times 10^{-28}$~g), $c$ is the light velocity (= $2.9979246 \times 10^{10}$~cm\,s$^{-1}$), and $k$ is the Boltzmann constant (= $1.3806488 \times 10^{-16}$~erg\,K$^{-1}$).
We take the values of $g_{i}, f_{ij}$ and $E_i$ from 
\citet{Kurucz92} (see also \citet{Barton+13} for SiO).
Also we take the values of $Q(T)$ for Na, K, {Fe,} Si and O from \citet{Irwin81}  and 
for SiO from \citet{Rossi+85}.

The function $H(\nu,a)$ is the line profile function given by \citep[e.g.,][]{Rybicki+86}
\begin{equation} 
	H(\nu,a) = \frac{V (\nu, a)}{\sqrt{\pi}\Delta \nu_D},
\end{equation}
where $a$ is the line width parameter, $V (\nu, a)$ is the Voigt function, and $\Delta \nu_D$ is the Doppler width given by
\begin{equation}
\Delta \nu_D=\frac{\nu_0}{c}\sqrt{\frac{2kT}{m_A}}, 
\label{eq:dw}
\end{equation}
where $m_A$ is the mass of the absorber and $\nu_0$ is the central frequency of the spectral line.
The line width parameter $a$ is calculated with damping constants for natural broadening and van der Waals broadening as \citep{Piskunov+01}
\begin{equation}
	a=\frac{\Gamma_r+\Gamma_{W}}{4\pi\Delta \nu_D}. 
\label{eq:a}
\end{equation}
{According to} \citet{Unsold55} and \citet{Gray76}, $\Gamma_r$ and $\Gamma_{W}$ are given by 
\begin{equation}
	\Gamma_r=0.222 \times 10^{14}\lambda^{-2}, 
\label{eq:gr}
\end{equation}
\begin{align}
	\Gamma_{W}=17 & \times 
	\left( 0.3 \times10^{-30} 
		\left[\frac{(l+1)^2}{(I-\chi_j)^2}-\frac{(l+1)^2}{(I-\chi_i)^2}\right] 
	\right)^{2/5} \notag \\
	& \times v^{3/5} N, 
\label{eq:gw}
	\end{align}
where $\lambda$ is wavelength in nm, 
$N$ is the total number density of the gas species in cm$^{-3}$, 
$I$ is the ionization potential in eV, 
$\chi_i$ is the excitation potential at the $i$th level in eV, 
and $l$ is the electric charge. 
$v$ is the relative thermal velocity defined by 
\begin{equation}
	v=v^A\sqrt{1+\frac{\mu^A}{\overline{\mu}}},
\label{eq:v}
\end{equation}
where $v^A$ is the thermal velocity of the absorber and $\mu^A$ and $\overline{\mu}$ are the molecular weight of the absorber and the mean molecular weight, respectively.
As for the SiO absorption opacity,
we assume $\Gamma_{W}=10^{-7}\times N$, following \citet{Kurucz+1981}. 
To calculate the line profile $H(\nu,a)$ numerically, we use a polynomial expansion of the Voigt function given by \citet{Kuntz97}. 
We have confirmed that this approximation yields relative errors of the order of $10^{-4}$,
 which is small enough for our discussion in this study.  

Figure~\ref{fig:cross} shows the cross section of each species multiplied by each molar fraction as a function of wavelength for $T =$ 2000~K(a) and 3000~K(b), respectively. 
The assumed pressures are $1.0\times10^{-5}$~bar for $T$~=~2000~K and 0.01~bar for $T$~=~3000~K.
The magma composition is assumed to be BSE. 
The calculated molar fractions of Na, K, {Fe,} Si, SiO, O, and O$_2$ are, respectively,  
$x_\mathrm{Na} 	= 0.721$, 
$x_\mathrm{K} 		= 0.0407$, 
{
$x_\mathrm{Fe} 	= 0.001$,
}
$x_\mathrm{Si} 	= 10^{-6}$, 
$x_\mathrm{SiO} 	= 0.00731$,
$x_\mathrm{O} 		= 0.0233$, and 
$x_\mathrm{O_2} 	= 0.189$ for $T =$ 2000~K(a) 
and 
$x_\mathrm{Na} 	= 0.204$, 
$x_\mathrm{K} 		= 0.0142$, 
{
$x_\mathrm{Fe} 	= 0.01$,
}
$x_\mathrm{Si} 	= 1 \times 10^{-3}$,
$x_\mathrm{SiO} 	= 0.343$, 
$x_\mathrm{O} 		= 0.0991$, and 
$x_\mathrm{O_2} 	= 0.211$ for $T =$ 3000~K(b).

In Fig.~\ref{fig:cross}a, 
{
the most abundant species Na (blue) presents a few hundred of strong absorption lines
  between 0.253 and 10.0~$\mathrm{\mu m}$, including the D line at 0.5893~$\mathrm{\mu m}$. 
Also, several hundred strong lines of K (magenta) that are distinguishable from the Na lines are also found between 0.286 and 10.0~$\mathrm{\mu m}$, including potassium doublet lines at 0.766 and 0.770 $\mathrm{\mu m}$. 
}
While SiO (red) is less abundant by a factor of about 100 than Na, {millions of} its electron-transition absorption lines dominate in the FUV wavelength region (0.14 - 0.3 $\mathrm{\mu m}$) and its absorption due to rotation-vibration transition dominates in several parts of the IR wavelength region{, except narrow line peaks of Na, K and Fe. }
The O electron-transition absorption (cyan) dominates between 0.10 and 0.13 $\mathrm{\mu m}$. 
 {
The several thousands lines of Fe (grey) dominates between 0.20 and 0.54 $\mathrm{\mu m}$. 
}
Although O$_2$ (green) is the second most {abundant} species, its absorption is obscured by the other absorptions except at a narrow range around 30~$\mathrm{\mu m}$. 
Also, the Si electron-transition absorption (yellow) is obscured by the SiO absorption in both Figs.~\ref{fig:cross}a and \ref{fig:cross}b and
presents many prominent absorption lines between 0.1 and 0.3 $\mathrm{\mu m}$. 

In the higher-$T$ case shown in Fig.~\ref{fig:cross}b, the SiO feature is more prominent.  
The electron-transition absorption in the FUV region is larger by about two orders of magnitude than that in Fig.~\ref{fig:cross}a, because the molar fraction of SiO is larger by two orders of magnitude (see Fig.~\ref{fig:COMP_X}) and comparable with Na at 3000~K. 
The most remarkable difference in absorption feature between Fig.~\ref{fig:cross}a and Fig.~\ref{fig:cross}b is that the SiO rotation-vibration absorption dominates at all the wavelengths longer than about 3~$\mathrm{\mu m}$,
{ except several hundreds of narrow lines of Na and K.}
 The rises in the peak values of SiO are due to the increase in abundance, while the enhancement of pressure broadening also contributes to the rises in absorption between those peaks.

{
Other minor species might contribute to the opacity. As for
 Mg and P present in Fig.~\ref{fig:COMP_X}, we have confirmed that the absorption opacities of these species are negligibly small {relative to} the SiO absorption opacity in most of the wavelength region of interest, except for narrow ranges between 0.1 and 0.15 $\mu$m.
Absorption line data of the diatomic molecules other than SiO in Fig.~\ref{fig:COMP_X} are unavailable, to our knowledge.
Thus, we have ignored their contribution.
}

Finally, for integrating the thermal structure of the atmosphere in the next section, we use the total absorption opacity, $\kappa_{\nu}$, calculated as 
\begin{equation}
	\rho\kappa_{\nu}=N \sum_{\mathrm{A}} x_\mathrm{A} \sigma^\mathrm{A}_{\nu}(i,j),
	\label{eq:kp}
\end{equation}
where 
$\rho$ is the mass density.

%%%
%%% SECTION 4
%%%
%\fi0
\section{TEMPERATURE PROFILE} \label{sec:structure}
In this study, we simulate the one-dimensional plane-parallel thermal structure of the atmosphere that is in the radiative, hydrostatic, and chemical equilibrium. 
To do so, we integrate the so-called two-stream equations with the assumption of quasi-isotropic radiation, adopting the $\delta-$Eddington approximation \citep{Toon+89}:
\begin{eqnarray}
	\frac{dF^+_{\nu}}{d\tau_{\nu}} 
		&=& \frac{7}{4}F^+_{\nu} + \frac{1}{4}F^-_{\nu} - 2\pi B_{\nu}(T),
		\label{eq:toon1} \\
	\frac{dF^-_{\nu}}{d\tau_{\nu}} 
		&=& -\frac{1}{4}F^+_{\nu} - \frac{7}{4}F^-_{\nu} + 2\pi B_{\nu}(T),
\label{eq:toon2}
\end{eqnarray}
where $F^+_{\nu}$ and $F^-_{\nu}$ are the upward and downward radiative fluxes, respectively, $B_\nu (T)$ is the Planck function, and $\tau_{\nu}$ is the optical depth defined by 
\begin{equation}
	\frac{d\tau_\nu}{dz}=-\kappa_\nu\rho,
\label{eq:tau0}
\end{equation}
where $z$ is the altitude. In equations~(\ref{eq:toon1}) and (\ref{eq:toon2}), 
we have assumed that the single-scattering albedo is 0, because absorption dominates scattering in most of the wavelength region of interest, as described above. 
{The boundary conditions at the top and bottom of the atmosphere are respectively
 $F^-_\nu =  0$ and  $F^+_\nu = \pi B_{\nu}(T_\mathrm{g})$, where $T_\mathrm{g}$ is the ground temperature.} 
{To solve equations~(\ref{eq:toon1}) and (\ref{eq:toon2}), 
we have used the widely-used algorithm developed by \citet{Toon+89}. 
According to \citet{Toon+89}, the error in the radiative fluxes caused by the two-stream approximation was typically less than 10\%, which is small relative to uncertainties in physical properties of exoplanets.}

The net radiative flux integrated over {all frequencies} is constant throughout the atmosphere in radiative equilibrium. 
For an irradiated atmosphere, 
the net radiative flux, $F_{\mathrm{net},\nu}$, is written as 
\begin{equation}
	F_{\mathrm{net},\nu}(\tau_{\nu})  = 
		F^+_{\nu} (\tau_\nu) - F^-_{\nu}(\tau_\nu) 
		- \mu_\ast F^\ast_{\nu} e^{-\tau_{\nu}/\mu_\ast},
\label{eq:fnet}
\end{equation}
where $F_\nu^\ast$ is the incident stellar flux and $\mu_\ast$ is the cosine of the zenith angle of the incident stellar radiation. 
The radiative equilibrium condition is given as 
\begin{equation}
	\int^{\infty}_{0}F_{\mathrm{net},\nu} \, d\nu = F_0,
\label{eq:hh}
\end{equation}
where $F_0$ is the constant flux. 
An HRSE is subject to tidal heating. 
Its strength depends on orbital and internal properties of the HRSE \citep{Barnes+10}. 
The detailed consideration of the tidal heating effect is beyond the scope of this study
{(A brief discussion is made in section~\ref{subsec:tidal})}. 
In this study, we assume $F_0$ = 10~W/m$^2$ for numerical stability. 
For $F_0 < 10$~W/m$^2$, we have confirmed that choice of $F_0$ have little influence on the thermal structure of the atmosphere.

We consider a pressure range from $1 \times 10^{-8}$ bar to the pressure at ground level, $P_\mathrm{g}$.  
Note that $P_\mathrm{g}$ is determined a posteriori, because $P_\mathrm{g}$ is the vapor pressure, which depends on the ground temperature, $T_\mathrm{g}$. 
Numerically, the atmosphere is vertically divided into 50 layers. 
The layers are prepared so that the size of each layer logarithmically increases with pressure. 
In addition, to integrate equations~(\ref{eq:toon1}) and (\ref{eq:toon2}), we consider 100 spectral intervals in a frequency range between $1\times$10$^{13}$ and 2$\times$10$^{15}$ s$^{-1}$ such that the size of each interval logarithmically increases with frequency. 
We calculate the harmonic mean opacities using cross sections obtained in section~\ref{sec:opacity}. 
The calculation method of the mean opacities is described in Appendix \ref{sec:cal_mo}.

To find the thermal structure of the atmosphere in radiative equilibrium and also in gas-melt chemical equilibrium with the underlying magma ocean, 
we need three iterative procedures. 
Firstly, to find the radiative-equilibrium {solution} 
{
for given opacities and molar fractions of atmospheric gases,
}
 we integrate the equation of energy conservation,  
\begin{equation}
	 \frac{dT_n}{d\widetilde{t}}=\frac{dF_n}{dP_n},
\label{eq:te}
\end{equation}
where $\widetilde{t}$ is a normalized time, $T_n$, $P_n$ and $F_n$ are the temperature, pressure, and net radiative flux in the $n$th layer, respectively.
The integration is continued until $\left|F_n-F_0\right|/F_0$ becomes smaller than 1~\% in all the layers.
 
Secondly, 
{
once we find
}
 the radiative-equilibrium solution, 
we calculate the opacities {that are consistent with temperatures obtained above.
In practice, to save memory and cpu time, we have prepared a numerical table in which the 
harmonic mean opacities (see Appendix \ref{sec:cal_mo}) in 100 spectral intervals} are given as functions of $T$, $\log{P}$, and, the mean molecular weight, $\overline{\mu}$: 
The grids are prepared for $T$ = 2000, 2500, 3000, 3500, 4000, and 4500~K, $\log (P/\mathrm{bar})$ = $-1$, $-2$, $-3$, $-4$, and $-5$, and $\overline{\mu}$ = 25, 30, and 35.
{In the pressure range from $1 \times 10^{-8}$~bar to $1 \times 10^{-5}$~bar, we have ignored the pressure dependence of the opacity because $\Gamma_r > \Gamma_{W}$ in the range.}
Note that we have checked that the calculated flux differs by at most 5~\%, even if a smaller table with $T$ = 2000, 3000, and 4000~K, $\log (P/\mathrm{bar})$ = $-1$, $-3$, and $-5$, and $\overline{\mu}$ = 25 and 35 is used. 
The convergence condition is that the temperature in each layer does not vary within less than 1~\%.

 \begin{figure}[htbp]
   \begin{center}
  \includegraphics[width=80mm]{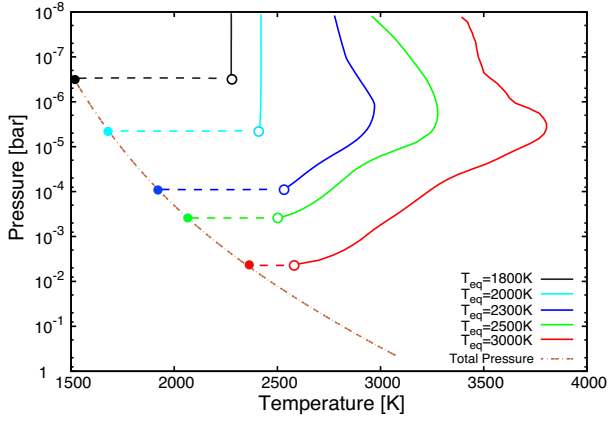}
   \end{center}
   \caption{
  Temperature-pressure profile in the mineral atmosphere on top of the BSE magma 
   of a super-Earth with gravity of 25~m/s$^2$
   for five choices of the substellar-point equilibrium temperature, $T_\mathrm{eq} $, 
   of 1800~K (black), 2000~K (cyan), 2300~K (blue), 2500~K (green), 
   and 3000~K (red) (see eq.~[\ref{eq:teq}]; $A_p = 0$). 
   The host star is assumed to be a Sun-like star with radius of 1~$R_\odot$ and 
   effective temperature of 6000~K and emit the blackbody radiation of 6000~K.    
    The solid lines show the day-side averaged profiles (i.e., the cosine of the stellar-light zenith angle $\mu_*=1/2$; see eq.~[\ref{eq:fnet}]). 
    The open circles show the temperatures at the bottom of the atmosphere, $T_\mathrm{b}$,
     and the filled circles show the temperatures at the ground, $T_\mathrm{g}$.
    The orange dotted line represents the total vapor pressure for the BSE composition, 
    which corresponds to the ground. }
\label{fig:HSE_TP_toon}
\end{figure}

 \begin{figure}[htbp]
 \begin{minipage}{0.5\hsize}
  \begin{center}
 (a) $T_\mathrm{eq}$ = 1800K
   \includegraphics[ width=80mm, clip]{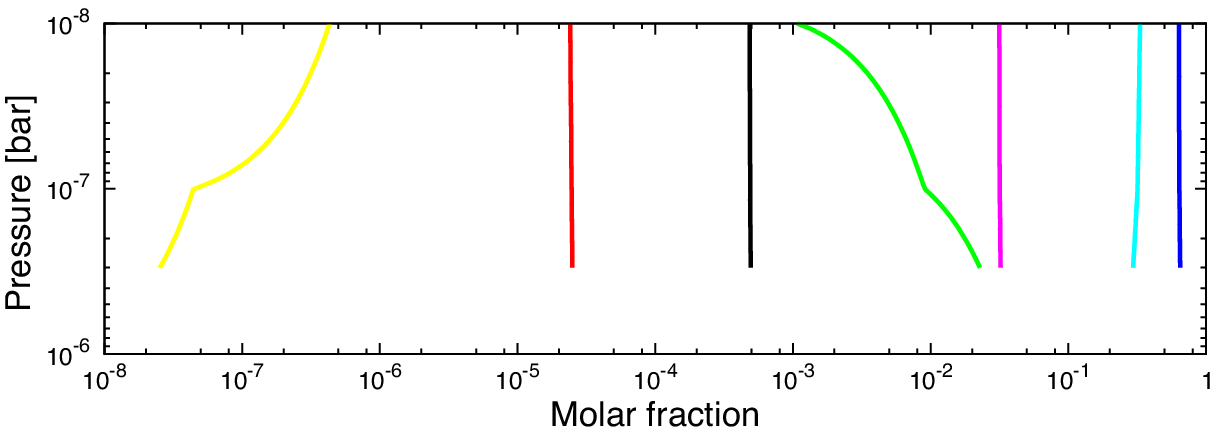}
  \end{center}
  \label{fig:mol18}
 \end{minipage}
 \begin{minipage}{0.5\hsize}
  \begin{center}
   (b) $T_\mathrm{eq}$ = 2000K
   \includegraphics[ width=80mm, clip]{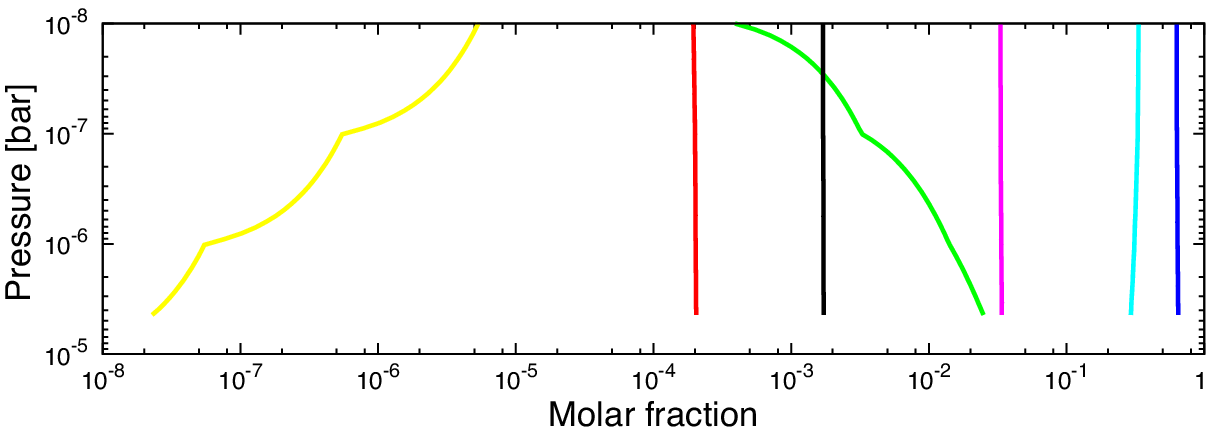}
  \end{center}
  \label{fig:mol20}
 \end{minipage} 
  \begin{minipage}{0.5\hsize}
  \begin{center}
    (c) $T_\mathrm{eq}$ = 2300K
   \includegraphics[ width=80mm, clip]{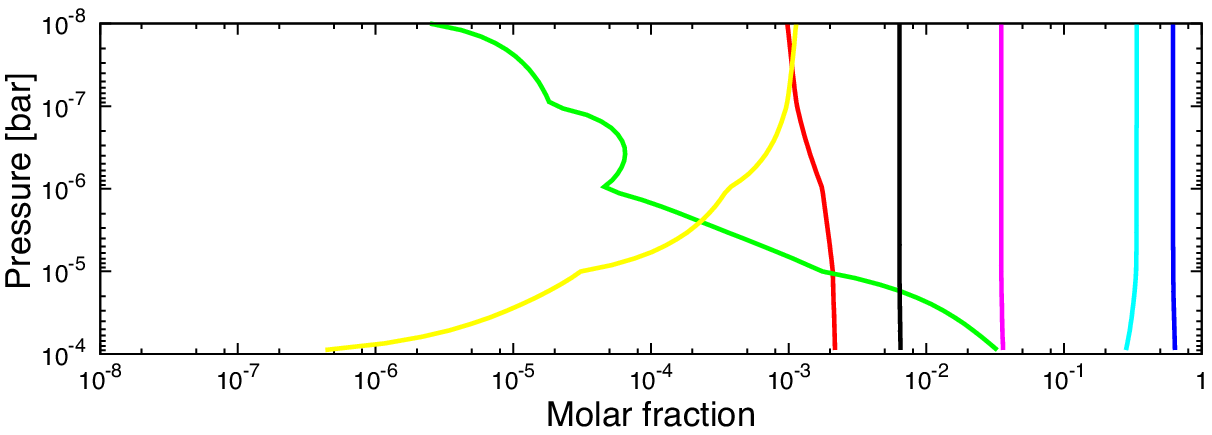}
  \end{center}
  \label{fig:mol23}
 \end{minipage}
 \begin{minipage}{0.5\hsize}
  \begin{center}
   (d) $T_\mathrm{eq}$ = 2500K
   \includegraphics[ width=80mm, clip]{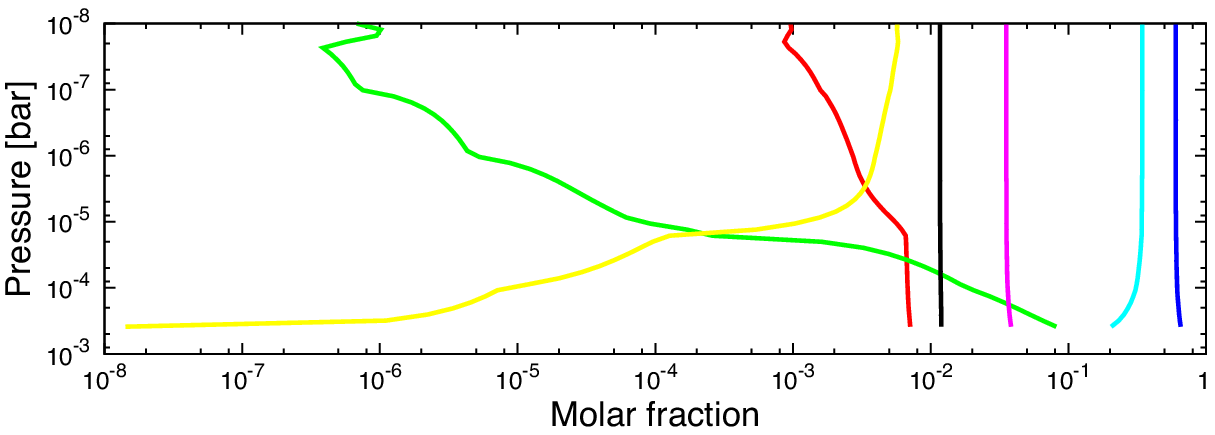}
  \end{center}
  \label{fig:mol25}
 \end{minipage}
  \begin{minipage}{0.5\hsize}
  \begin{center}
    (e) $T_\mathrm{eq}$ = 3000K
   \includegraphics[width=80mm, clip]{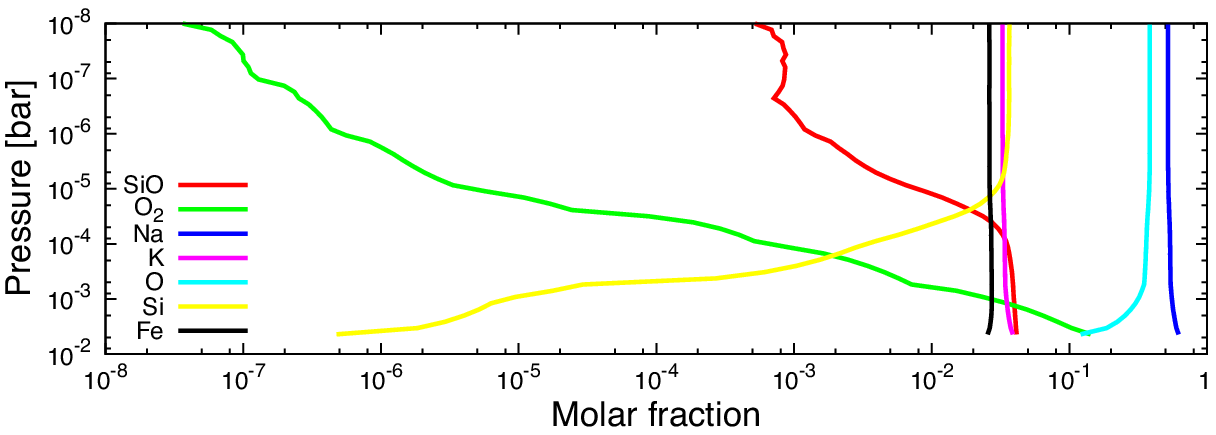}
  \end{center}
  \label{fig:mol30}
 \end{minipage}
  \caption{
    The distribution of the seven major gas species in chemical equilibrium in the mineral atmosphere. 
  Molar fractions of  SiO (red), 
    O$_2$ (green), Na (blue), K (magenta), Fe (grey), O (cyan) and Si(yellow) are shown as functions of pressure
      for five choices of the substellar-point equilibrium temperature;  (a) $T_\mathrm{eq}$ = 1800K, 
  (b) 2000K, 
  (c) 2300K, 
  (d) 2500K, and
  (e) 3000K. 
  The corresponding temperature-pressure structure of the atmosphere is shown in Figure \ref{fig:HSE_TP_toon}.
}
 \label{fig:moltp}
\end{figure}

Finally, we have to find the solution in which the atmosphere is in equilibrium with the underlying magma ocean at its bottom. The ground pressure $P_\mathrm{g}$ and molar fractions $x_\mathrm{A}$ are functions of the ground temperature $T_\mathrm{g}$ (see section~\ref{sec:chemistry}). 
From equations~(\ref{eq:fnet}) and (\ref{eq:hh}), the magma being assumed to be a blackbody, $T_\mathrm{g}$ is given as 
\begin{equation}
	\sigma T_\mathrm{g}^4 = 
		F_0 + 
		\int_0^\infty \left\{ F^-_{n,\nu}(P_\mathrm{g}) 
		+ \mu_\ast F^\ast_{\nu} e^{-\tau_{\nu, \mathrm{g}}/\mu_\ast} \right\} d\nu,
\label{eq:tg_ite}
\end{equation}
where $\sigma$ is the Stefan-Boltzmann constant ($=5.67\times10^{-8}$W/m$^2$K$^4$) and $\tau_{\nu,\mathrm{g}}$ is the total optical depth of the atmosphere.
We calculate $T_\mathrm{g}$ from Eq.~(\ref{eq:tg_ite}) and then calculate $P_\mathrm{g}$ and $x_\mathrm{A}$.
The gas chemical equilibrium composition in each layer, $x_\mathrm{A}$, is obtained by interpolating linearly the tabulated data made by the NASA CEA code.
The grids are prepared for
 $T$ = 2000, 2500, 3000, 3500, 4000, and 4500~K,
  $\log (P/\mathrm{bar})$ = $-1$, $-2$, $-3$, $-4$, $-5$, $-6$, $-7$ and $-8$,
   and $T_\mathrm{g}$ being every one kelvin from 1500K to 3000K.  
The convergence condition is that $T_\mathrm{g}$ and the temperature in each layer do not vary within less than 1~\%.

Figure~\ref{fig:HSE_TP_toon} shows the calculated temperature profiles for different substellar-point equilibrium temperatures defined by 
\begin{equation}
	T_\mathrm{eq}^4 = (1 - A_p) \frac{R_\ast^2}{D^2} T_\ast^4,
\label{eq:teq}
\end{equation}
where $R_\ast$ and $T_\ast$ are respectively the radius and temperature of the host star, $A_p$ is the planetary albedo, and $D$ is the orbital distance of the planet. 
Five values of $T_\mathrm{eq}$ are chosen: 
$T_\mathrm{eq} $ = 1800~K (black), 2000~K (cyan), 2300~K (blue), 2500~K (green), and 3000~K (red). 
The incident stellar flux is calculated as $F^\ast_{\nu}=T_\mathrm{eq}^4/T_\ast^4 \times \pi B_\nu(T_\ast)$.
{
The solid lines represent the temperature profiles at $\mu_*=1/2$ (i.e., day-side average).
Also, the open and filled circles show the temperatures at the bottom of the atmosphere, $T_\mathrm{b}$, and at the ground, $T_\mathrm{g}$, respectively.
}
Figure~\ref{fig:moltp} shows the calculated molar fraction profiles of {seven} major species for the atmosphere structure shown in Fig.~\ref{fig:HSE_TP_toon}.
Here we have assumed $g$ = 25~m/s$^2$ and $A_p$ = 0. The host star is assumed to emit blackbody radiation of 6000~K. The magma composition is assumed to be BSE.

The most remarkable feature of the temperature profiles in Fig.~\ref{fig:HSE_TP_toon}
 is the thermal inversion structure in the cases of $T_\mathrm{eq} \geq 2300$~K.  
As found in section~\ref{sec:opacity}, 
the absorption is more significant in the UV and visible wavelength regions (0.1 to $\sim$1~$\mathrm{\mu m}$) than in the IR region ($\sim$1 to 4 $\mathrm{\mu m}$), 
mainly because of the strong UV-ray absorption by SiO and visible-light absorption by {the doublet lines of} Na and K. 
{
(Note that most of the absorption lines of Fe, Na and K are too narrow to contribute to the mean opacity.)
}
This means that the absorption of the incident stellar radiation is stronger compared to the intrinsic planetary radiation,
 which results in thermal inversion for $P$ $\geq10^{-5}$bar.
 { 
For $P$ $\leq10^{-5}$bar, the temperature decreases with decreasing pressure,
 because the dissociation of SiO occurs as shown in Figure~\ref{fig:moltp}, so that 
 the UV-ray absorption by SiO becomes weak.
 }
 
 {
  Not only the thermal inversion region extends but  $T_\mathrm{g}$ increases, as $T_\mathrm{eq}$ increases.}
In the case of $T_\mathrm{eq}$ {=} 3000~K, $T_\mathrm{g}$, is significantly smaller than $\mu_*^{1/4}T_\mathrm{eq}$. 
In contrast, $T_\mathrm{g} \simeq \mu_\ast^{1/4}T_\mathrm{eq}$ for low $T_\mathrm{eq}$, because the atmosphere is so optically thin that the ground is directly heated by stellar irradiation.
In the cases of $T_\mathrm{eq} \leq 2000$~K, the atmosphere is thus isothermal. 
Note that jumps in temperature between $T_\mathrm{b}$ and $T_\mathrm{g}$ are found in Figure \ref{fig:HSE_TP_toon}. 
This is because those atmospheres are optical thin for the incident stellar flux and any heat transfer process except radiative transfer is not taken into account.
{
Thus, in reality, the atmosphere is not exactly isothermal,
 because there should be a conductive region with steep temperature gradient near the ground.
 }
 Such structure, however, has little impact on the emission spectra that we use when discussing the detectability of HSREs in secondary eclipse (see section~\ref{sec:detectability}). 

%%%
%%% SECTION 5
%%%
\section{DETECTABILITY VIA SECONDARY ECLIPSE OBSERVATION} \label{sec:detectability}
%
% Section 5.1
%
\subsection{Secondary eclipse depth}

Here we assess the detectability of the mineral atmospheres of HRSEs that we have modeled via the secondary eclipse observation. 
This observation enables us to infer the vertical temperature profile and composition of an exoplanetary atmosphere by obtaining planetary emission spectrum. 
In particular, the thermal inversion structure found in section~\ref{sec:structure} is expected to be detected. 

The secondary eclipse observation measures the ratio of the planetary luminosity 
to the sum of  the stellar and planetary luminosities. 
The ratio is often called the secondary eclipse depth, $\epsilon_\lambda$, which is given by 
\begin{equation}
	\epsilon_{\lambda} \simeq
	\left( \frac{R_p}{R_*} \right)^2
	\frac{F_{p, \lambda}+A_p F_{*, \lambda}}{F_{*, \lambda}},
\label{eq:dp}
\end{equation}
where $F_{\ast, \lambda}$ is the emergent stellar flux and $F_{p, \lambda}$ is the emergent planetary flux calculated as 
\begin{equation}
	F_{p, \lambda}=\pi\int^{\infty}_0B_{\lambda}(T(\tau_{\lambda}))e^{-\tau_{\lambda}}d\tau_{\lambda}.
\label{eq:fp}
\end{equation}
Below, as in section~\ref{sec:structure}, we assume $g = 25$~m/s$^2$, $A_p = 0$, and blackbody stellar radiation of 6000~K.
 The magma composition is assumed to be BSE. Also, we suppose an HRSE with radii of 2~$R_\oplus$ orbiting a solar analog, namely $R_p / R_\ast$ = 0.02. 
 To integrate equations~(\ref{eq:dp}) and (\ref{eq:fp}),
  we consider 30000 spectral intervals in the wavelength range between $0.1$ and $100$ $\mathrm{\mu m}$;
    the size of each interval logarithmically increases with wavelength. 
   Then, as in section~\ref{sec:structure}, we calculate the arithmetic mean opacities, interpolating linearly the tabulated data. 
We consider only the case of $\mu_\ast = 1/2$, because we are interested in planetary disk averaged emission (i.e., day-side average). 

\begin{figure}[tbp]
  \begin{minipage}{1.0\hsize}
  \begin{center}
   \includegraphics[width=1.0\linewidth]{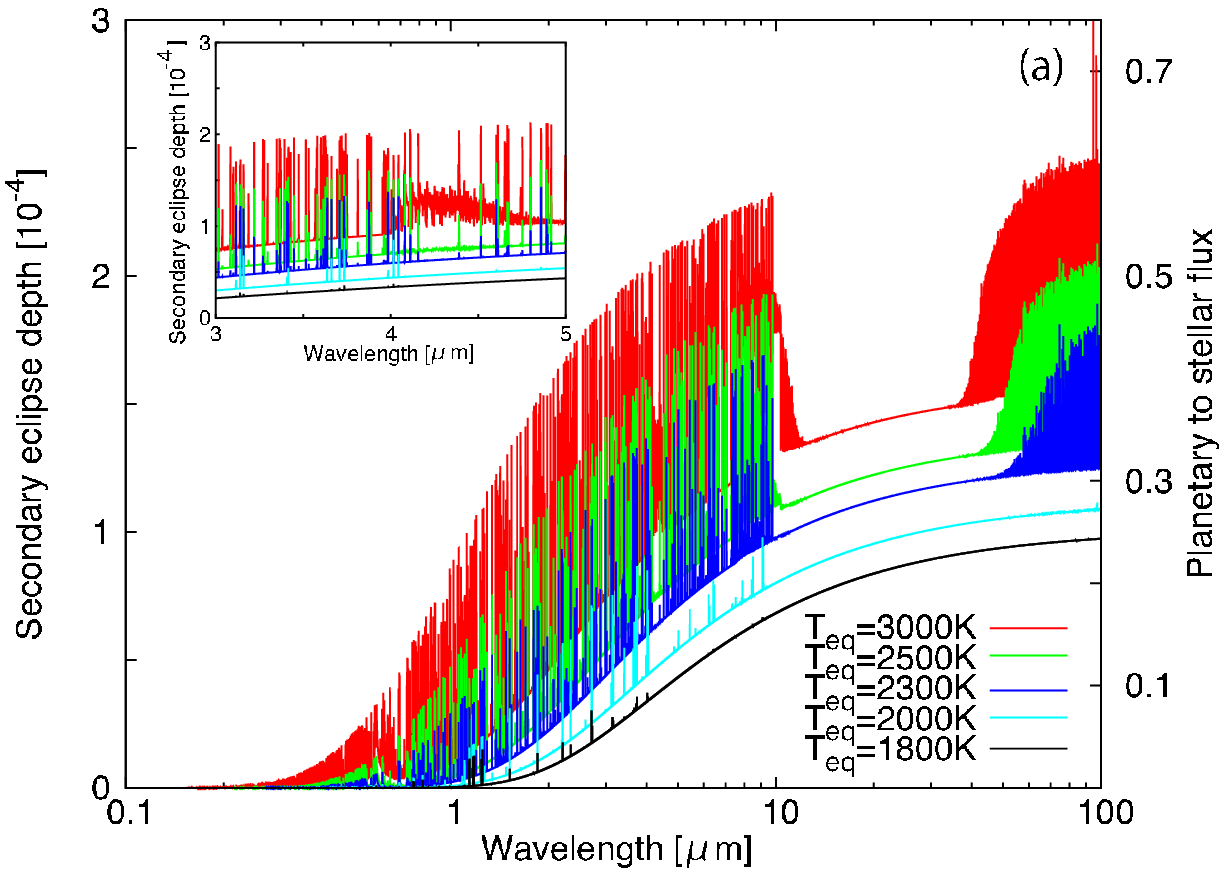}
  \end{center}
  \label{fig:dep_day}
 \end{minipage}
 \begin{minipage}{1.0\hsize}
  \begin{center}
   \includegraphics[width=1.0\linewidth]{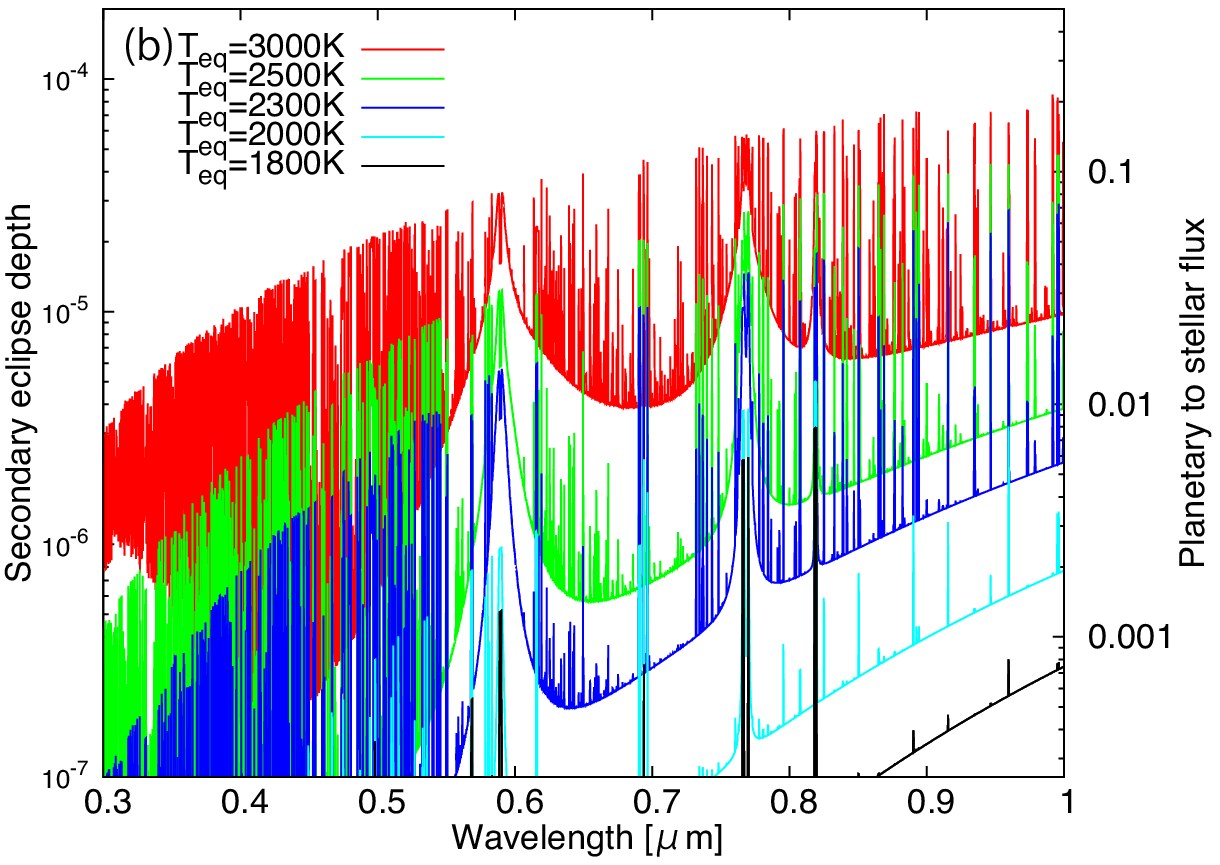}
  \end{center}
  \label{fig:dep_day_vis}
 \end{minipage}
   \caption{
   Predicted dayside-averaged emission spectra of a hot rocky super-Earth of 2~$R_\oplus$ 
   in secondary eclipse that has a mineral atmosphere in equilibrium 
   with an underlying BSE magma ocean. 
   The secondary eclipse depth (see eq.~[\ref{eq:dp}]) is shown as a function of 
   wavelength in the range of (a) 0.1--100~$\mathrm{\mu m}$ and (b) 0.3--1~$\mathrm{\mu m}$. 
   {The inset in the Fig.(a) is an enlarged view of the spectra in 3--5~$\mathrm{\mu m}$.}
   Five equilibrium temperatures are chosen: $T_\mathrm{eq} $ = 1800~K (black), 
   2000~K (cyan), 2300~K (blue), 2500~K (green), and 3000~K (red). 
   The corresponding atmospheric structures are shown by dashed lines 
   in Fig.~\ref{fig:HSE_TP_toon}. 
   The host star is assumed to be a Sun-like star with radius of 1~$R_\odot$ 
   (i.e., the planetary/stellar radius ratio being 0.02) and emit the blackbody radiation of 6000~K. 
   }
 \label{fig:HSE_dep}
\end{figure}

Figure~\ref{fig:HSE_dep} plots the secondary eclipse depth $\epsilon_{\lambda}$ (and also $F_{p,\lambda}/F_{\ast, \lambda}$) for $T_\mathrm{eq}$ = 1800~K (black), 2000~K (cyan), 2300~K (blue), 2500~K (green), and 3000~K (red): Panels (a) and (b) show the spectra in the wavelength region of 0.1--100 $\mathrm{\mu m}$ and 0.3--1 $\mathrm{\mu m}$, respectively. 
The secondary-eclipse-depth spectra for $T_\mathrm{eq}=1800$~K and $2000$~K 
{
exhibit only some narrow line features of Na and K.
This is mainly because the atmospheres are isothermal (see Fig.~\ref{fig:HSE_TP_toon})
 except for the temperature jump above the ground. }
 {
The line features are, thus, due to the emission from {parts of} the atmosphere with temperatures higher than $T_\mathrm{g}$, while the continuous spectra are the black body spectra of $T_\mathrm{g}$.
}
 
By contrast, the spectra for $T_\mathrm{eq}=2300$~K {and $2500$~K} present noticeable features of SiO around 10 and 100 $\mathrm{\mu m}$,
in addition to hundreds of narrow features of Na and K.
{
For $T_\mathrm{eq}=3000$~K, a spectral feature by SiO around 4~$\mathrm{\mu m}$ appears in addition to those around 10 and 100~$\mathrm{\mu m}$, {as shown in the subfigure inside Fig.\ref{fig:HSE_dep}a}. 
}
The features of SiO become stronger as $T_\mathrm{eq}$ increases, because the SiO abundance is sensitive to temperature and also the thermal inversion is more significant at higher temperature. 
{
Additionally, one can also identify features of the sodium doublet at 0.589 $\mathrm{\mu m}$ and the potassium doublet at 0.77 $\mathrm{\mu m}$ and narrow line features of Fe, Na and K in the visible band, as shown in Panel~(b).
}

%
% Section 5.2
%
\subsection{Detectability}
\label{subsec:dep_r2}
We evaluate the detectability of the mineral atmosphere of an HRSE for near-future space-based secondary-eclipse observation. The photometric accuracy of the transit/eclipse observation from space depends on the instrument measurement noise, the intrinsic stellar variability, and the shot noise from photon statistics \citep[e.g.][]{2012PASP..124..700C}. In this section, we quantify the photometric accuracy required for the line detection and discuss the feasibility, assuming the ideal case that the total noise is dominated by the shot noise (i.e. the photon-noise limit).
To quantify the detectability for the photon-noise limit, we introduce the signal-to-noise ratio of the detection of the secondary eclipse as 
\begin{eqnarray}
(S/N)_{\Delta \lambda} &=& \frac{N_p}{\sqrt{N_*}}, \\
&\simeq& \epsilon_{\lambda, \Delta \lambda} \sqrt{\frac{At_\mathrm{obs}}{\mbox{\Pisymbol{psy}{"C2}}}\frac{R_{*}^2}{d^2}\frac{\pi B_{\Delta\lambda}(T_*)}{h}},
\label{eq:snr}
\end{eqnarray}
where $\Delta \lambda$ is the band width used for the detection of the secondary eclipse,
 $N_*$ is the photon counts of the star,
 $A$ is the photon collecting area,
 $t_\mathrm{obs}$ is the exposure time,
 $d$ is the distance of the system,
 \Pisymbol{psy}{"C2} is the resolving power of the spectra,
 $\epsilon_{\lambda, \Delta \lambda}$ is the depth of the secondary eclipse within $\Delta \lambda$,  $N_p$ is the photon count of the planet, $B_{\Delta\lambda}(T_*)$ is the blackbody radiation integrated over {a  wavelength band} (i.e., $\int_{\Delta\lambda} B_{\lambda}(T_*)d\lambda/\Delta \lambda$), and $h$ is the Planck constant. 

\begin{figure}[bp]
\begin{tabular}{c}
 \begin{minipage}{0.950\hsize}
\begin{center}
\includegraphics[width=1.0\linewidth]{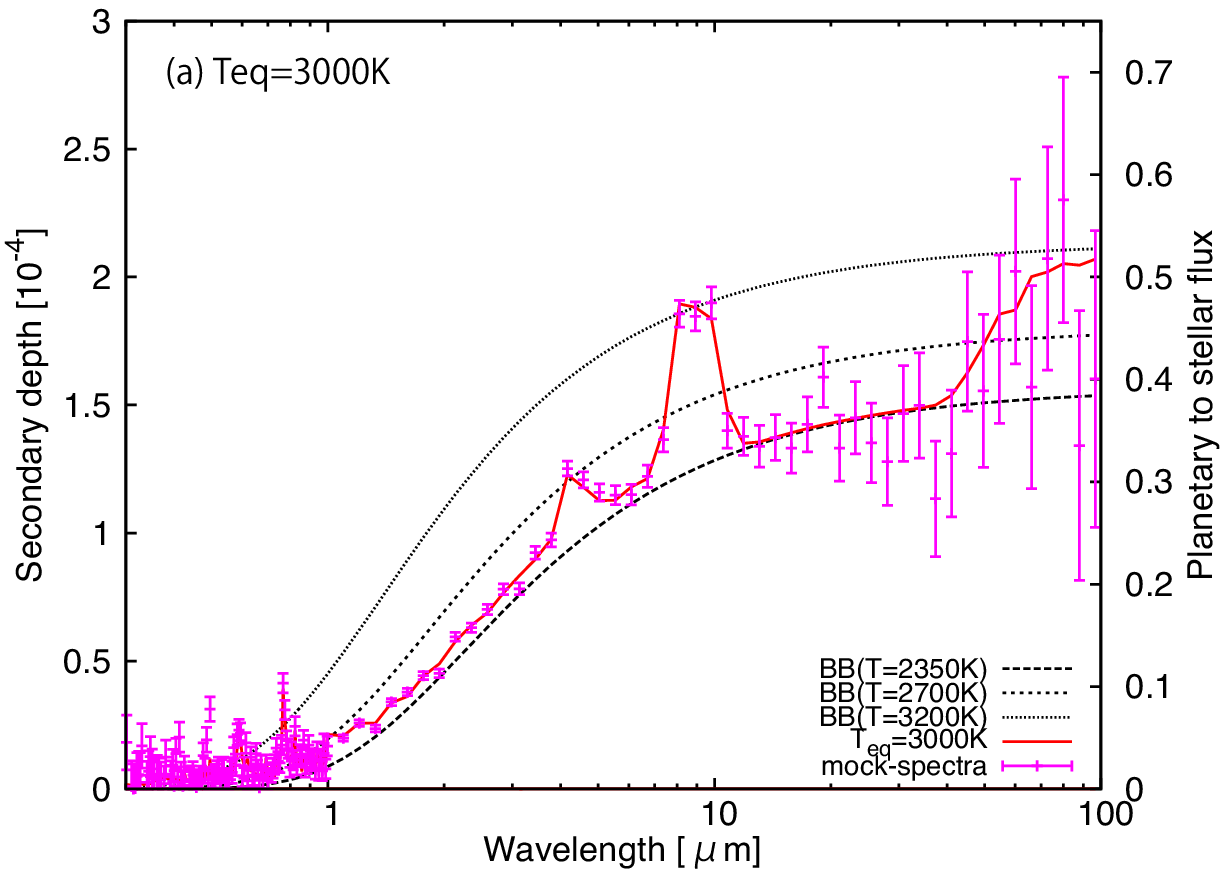}
\end{center}
 \end{minipage}
 \\
 \begin{minipage}{0.950\hsize}
\begin{center}
\includegraphics[width=1.0\linewidth]{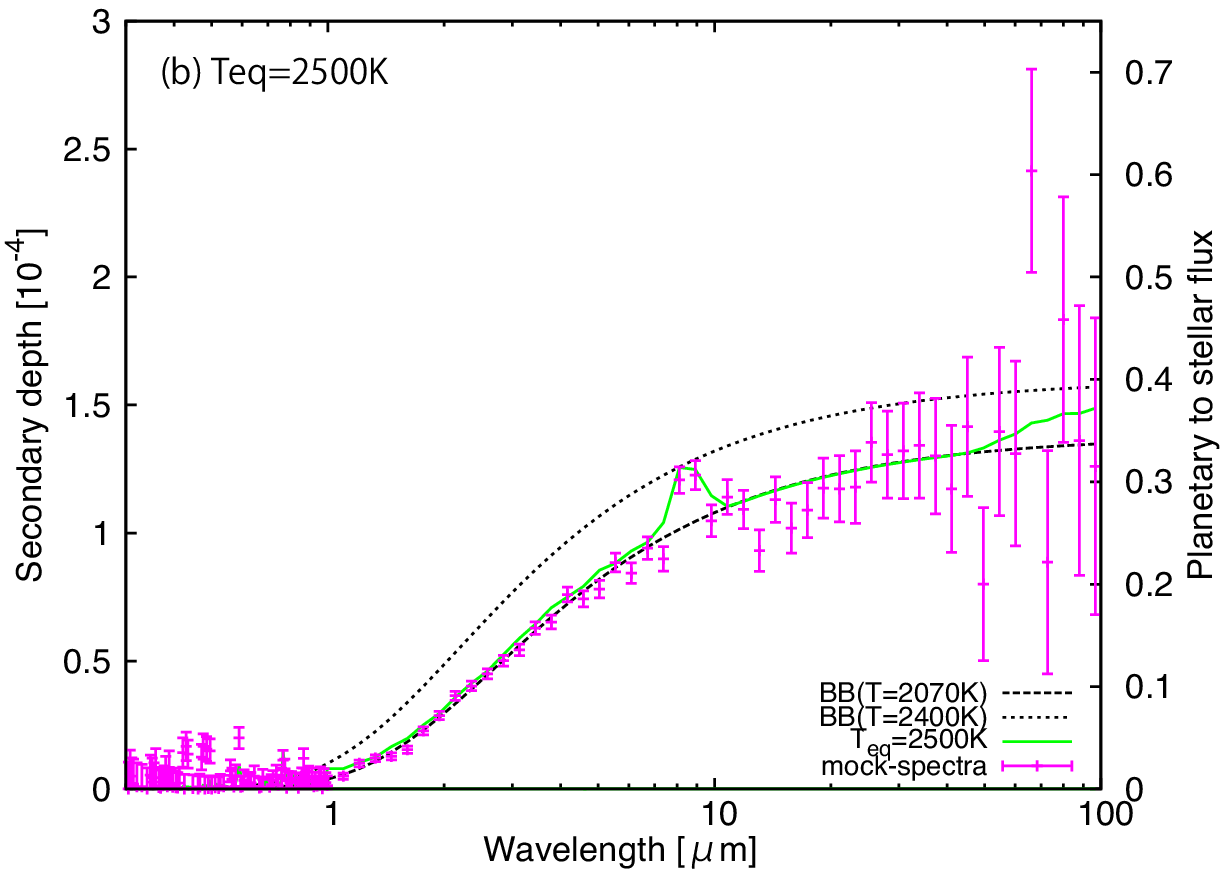}
\end{center}
 \end{minipage}
 \\
 \begin{minipage}{0.950\hsize}
\begin{center}
\includegraphics[width=1.0\linewidth]{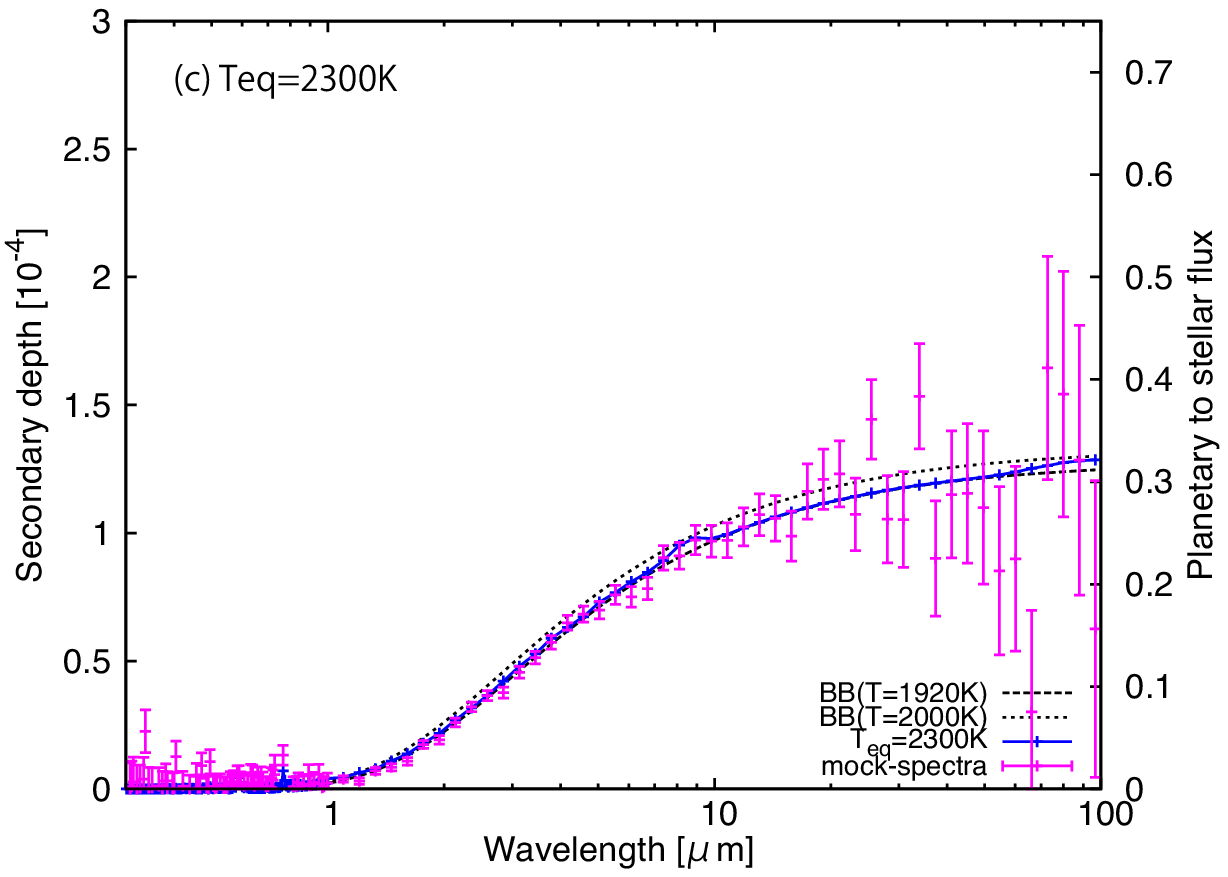}
\end{center}
 \end{minipage}
 \end{tabular}
  \caption{The secondary eclipse depths theoretically predicted and the mock 
  observational spectrum compared to the blackbody spectra (BB). 
  The secondary eclipse depths are shown as functions of wavelength for 
  $T_\mathrm{eq}$ = (a) 3000~K, (b) 2500~K, and (c) 2300~K. 
  The black dotted lines show the secondary eclipse depths of the blackbody, 
  temperatures of which are indicated by ''BB($T$)''. 
  The bars show the mock spectrum calculated based on
   the photon-noise limit with the assumptions that $R_*$ = 1~$R_{\odot}$, $R_p$ = $0.02 R_{\odot}$, $A=\pi (5 \mathrm{m}/ 2)^2$, $t_\mathrm{obs}$ = 10~hours, \Pisymbol{psy}{"C2} =  100 in 0.3-1 $\mathrm{\mu m}$ and 10 in 1-100 $\mathrm{\mu m}$, and $d$ = 100 pc (see the text for the details). 
  The corresponding atmospheric structures are shown by dashed lines 
   in Fig.~\ref{fig:HSE_TP_toon}. 
   }
 \label{fig:mockdep}
\end{figure}

Figure~\ref{fig:mockdep} plots the theoretical (solid line), mock (cross), and blackbody (dotted lines) secondary-eclipse-depth spectra for $T_\mathrm{eq}$ = 3000~K (a), 2500~K (b), and 2300~K (c). 
{
When calculating the mock spectra, we use the false normal random number generation circuit \citep{NRC+96}, and 
}
include the statistical noise, assuming the normal distribution with the standard deviation $\sigma = \sqrt{N_*}$.  We assume $A = (5/2 \, {\rm m})^2$ as a fiducial value, roughly corresponding to that of James Webb Space Telescope \citep[JWST;][]{Clampin09}, and that \Pisymbol{psy}{"C2} = 100 in 0.3-1 $\mathrm{\mu m}$ and 10 in 1-100 $\mathrm{\mu m}$, $t_\mathrm{obs}$ = 10~hours, and $d$ = 100~pc.
  The blackbody spectra are calculated from equation~(\ref{eq:dp}) with the assumption $F_{p,\lambda}$ = $B_{\lambda}(T)$. 

 As demonstrated in Fig.~\ref{fig:mockdep}, the spectral features of Na, K and SiO are represented by the blackbody spectra with different temperatures (i.e., brightness temperature $T_\mathrm{br}$). 
 {
For example, in the case of $T_\mathrm{eq}$ = 3000~K (see Fig.~\ref{fig:mockdep}a),
 $T_\mathrm{br}$ $\simeq$ 3200~K at 10$\mathrm{\mu m}$, and 2700~K at 4$\mathrm{\mu m}$ due to SiO.
 Additionally, although not shown,
$T_\mathrm{br}$ $\simeq$ 3500~K at 0.6~$\mathrm{\mu m}$ and at 0.8$\mathrm{\mu m}$ 
  (which are due to Na and K, respectively), and 3100~K at 100~$\mathrm{\mu m}$ (SiO). 
 }
At the other wavelengths, $T_\mathrm{br} \simeq T_\mathrm{g}$ (e.g., 
$T_\mathrm{br}$ = 2350~K for $T_\mathrm{eq}$ = 3000~K), because the atmosphere is optically thin at those wavelengths.

 \begin{figure}[htbp]
\begin{tabular}{c}
 \\
 \begin{minipage}{1.0\hsize}
\begin{center}
\includegraphics[width=0.950\linewidth]{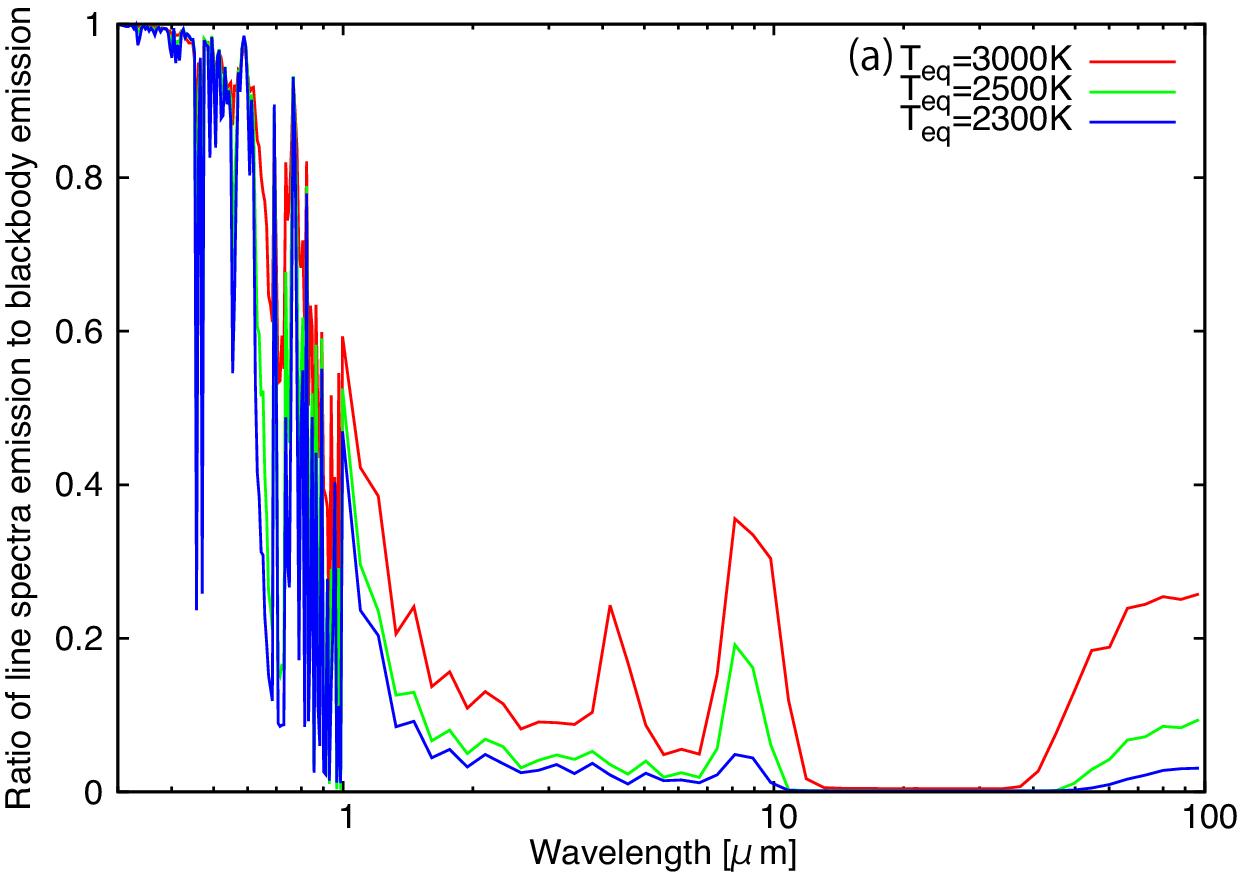}
\end{center}
 \end{minipage}
 \\
 \begin{minipage}{1.0\hsize}
\begin{center}
\includegraphics[width=0.950\linewidth]{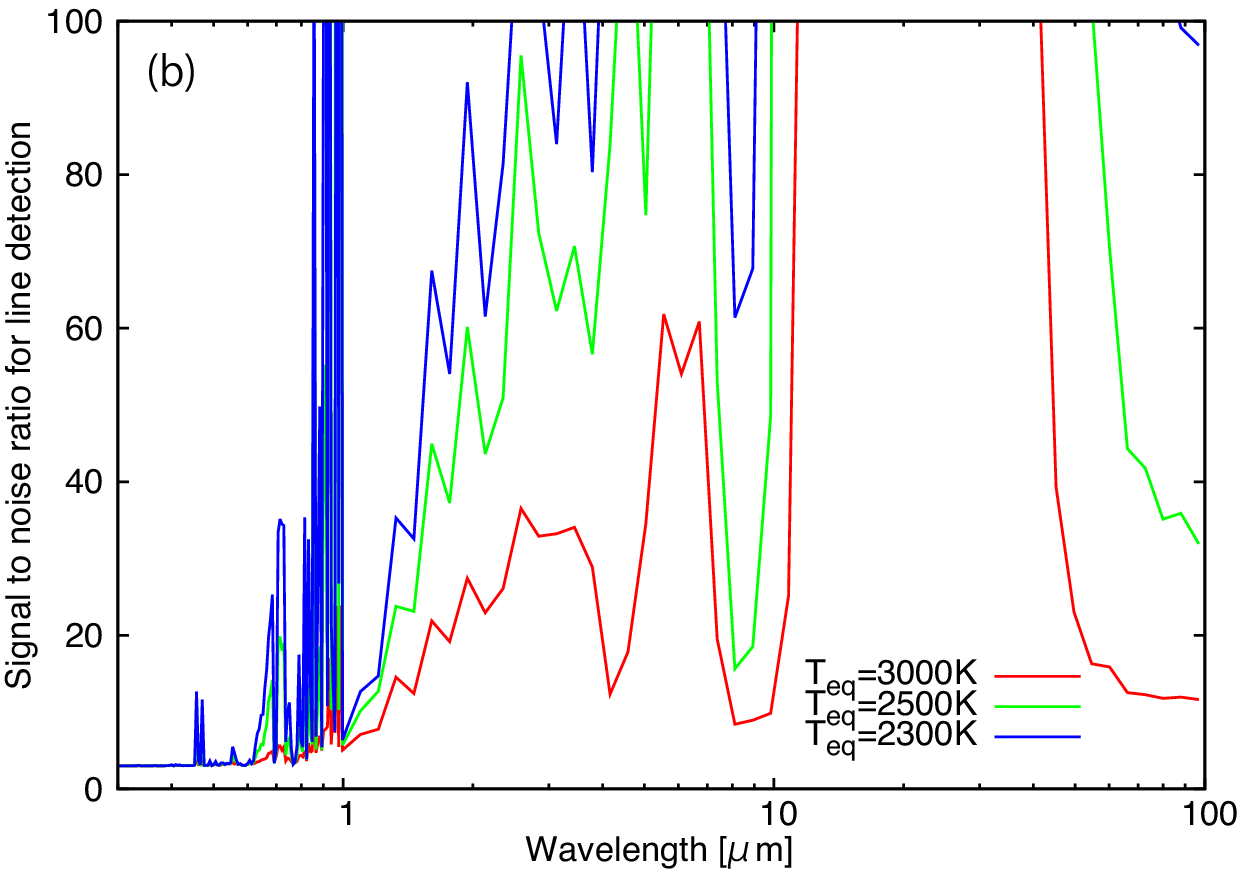}
\end{center}
 \end{minipage}
 \end{tabular}
   \caption{(a) The secondary eclipse depth relative to that of blackbody with $T_\mathrm{g}$ 
   ($\eta_{\lambda,\Delta \lambda}$; see eq.~[\ref{eq:eta}])
    and 
   (b) the minimum signal-to-noise ratio required to distinguish the emission 
   spectral features from the blackbody emission with $T_\mathrm{g}$
   ($|(S/N)_{\Delta \lambda}|_\mathrm{min}$; see eq.~[\ref{eq:snld}]).
    \Pisymbol{psy}{"C2} = 100 in 0.3-1 $\mathrm{\mu m}$ and 10 in 1-100 $\mathrm{\mu m}$. 
   Three equilibrium temperatures are chosen: $T_\mathrm{eq} $ = 3000~K (red), 
   2500~K (green), 2300~K (blue). 
   The corresponding atmospheric structures are shown by dashed lines 
   in Fig.~\ref{fig:HSE_TP_toon}. 
   The host star is assumed to be a Sun-like star with radius of 1~$R_\odot$ 
   (i.e., the planetary/stellar radius ratio being 0.02) and 
   emit the blackbody radiation of 6000~K. 
 }
 \label{fig:snl}
\end{figure}

Comparing the mock spectra with the blackbody spectra, one can quantify the detectability of the spectral features of Na, K and SiO. 
The error bars become larger in both shorter and longer wavelength regions, because the number of incident photons (i.e., $B_{\Delta\lambda}(T_*)$) is small in such wavelength regions. 
Also, the values of \Pisymbol{psy}{"C2}, which are assumed to be 100 in 0.1-1 $\mathrm{\mu m}$ and 10 in 1-100 $\mathrm{\mu m}$, affect the value of $(S/N)_{\Delta\lambda}$. 
As shown in Figure  ~\ref{fig:mockdep}a, the spectral features of SiO are sufficiently detectable at 4, 10 and 100~$\mathrm{\mu m}$ for $T_\mathrm{eq}$ = 3000~K. 
The SiO feature is also sufficiently detectable at 10~$\mathrm{\mu m}$ for $T_\mathrm{eq}$ = 2500~K (Fig.~\ref{fig:mockdep}b), while the feature is marginally undetectable for $T_\mathrm{eq}$ = 2300~K (Fig.~\ref{fig:mockdep}c).

The signal-to-noise ratio for detecting the line absorption spectral feature of the secondary eclipse depth is generally smaller than that for the secondary eclipse detection.
  Thus, we define the secondary eclipse depth relative to
   that with no atmospheric feature (i.e., blackbody with $T_\mathrm{g}$),
    $\eta_{\lambda,\Delta \lambda}$ as
      \begin{equation}
\eta_{\lambda,\Delta \lambda} \equiv \left|
\int^{\lambda+\Delta \lambda}_{\lambda}
\frac{\epsilon_{\lambda}-\epsilon_{\mathrm{BB},\lambda} (T_\mathrm{g})}{\epsilon_{\lambda}}
\frac{d\lambda}{\Delta \lambda}
 \right|,
\label{eq:eta}
\end{equation}
where $\epsilon_{\mathrm{BB,\lambda}}(T_\mathrm{g})$ is the secondary eclipse depth by the blackbody radiation with $T_\mathrm{g}$.
 Using $\eta_{\lambda,\Delta \lambda}$, we estimate the detectability of the line spectral features.
  From equations (\ref{eq:snr}) and (\ref{eq:eta}), 
  {
  we introduce a new quality, }
   $(S/N)_{L,{\Delta \lambda}}$, that is defined as
\begin{eqnarray}
(S/N)_{L,{\Delta \lambda}} &=& \eta_{\lambda,\Delta \lambda} (S/N)_{\Delta \lambda}.
\label{eq:snld}
\end{eqnarray}
 When $(S/N)_{L,{\Delta \lambda}}$ is sufficiently large, we can detect the line spectral feature of the secondary eclipse depth.
 {
 Thus, if the minimum value of $(S/N)_{L,{\Delta \lambda}}$ required to detect a line spectral feature,
 $|(S/N)_{L,{\Delta \lambda}}|_\mathrm{min}$, is 3,  
  $(S/N)_{\Delta \lambda} \geq 3/ \eta_{\lambda,\Delta \lambda}$ is required to do so.
  }

 Figure~\ref{fig:snl}a shows $\eta_{\lambda,\Delta \lambda}$ for $T_\mathrm{eq}$ = 3000~K (red), 2500~K (green), and 2300~K (blue)
 with \Pisymbol{psy}{"C2} = 100 in 0.3-1 $\mathrm{\mu m}$ and 10 in 1-100 $\rm{\mu}$m.
 As shown in Figure~\ref{fig:snl}a, 
  $\eta_{\lambda,\Delta \lambda}$ takes large values at some wavelengths,
   because of the strong line features:
    $\eta_{\lambda,\Delta \lambda} \sim$ 1 at 0.6 $\mathrm{\mu m}$ (Na) and 
    $\sim$ {0.35} at 10 $\mathrm{\mu m}$ (SiO) for $T_\mathrm{eq}$ = 3000~K (red). 
   {$\eta_{\lambda,\Delta \lambda}$ is large at $\sim$ 4 $\mathrm{\mu m}$
    because of the SiO feature peak only for $T_\mathrm{eq}$ = 3000~K (red),
    since the SiO feature is weaker than Na and K features for $T_\mathrm{eq}$ = 2300~K and 2500~K.}
   Also,
  Figure~\ref{fig:snl}b shows 
  $|(S/N)_{L,{\Delta \lambda}}|_\mathrm{min} / \eta_{\lambda,\Delta \lambda}
   (\equiv |(S/N)_{\Delta \lambda}|_\mathrm{min}$) for $T_\mathrm{eq}$ = 3000~K (red), 2500~K (green), and 2300~K (blue),
   assuming $|(S/N)_{L,{\Delta \lambda}}|_\mathrm{min}$ = 3 as a  minimum fiducial value to detect the line spectral feature with \Pisymbol{psy}{"C2} = 100 in 0.3-1 $\mathrm{\mu m}$ and 10 in 1-100 $\mathrm{\mu m}$.
 We obtain that $|(S/N)_{\Delta \lambda}|_\mathrm{min} \lesssim$ 100 around 4, 10 and 100~$\mathrm{\mu m}$;
 for example, $|(S/N)_{\Delta \lambda}|_\mathrm{min}$ $\simeq$  {60} at 10~$\mathrm{\mu m}$ for $T_\mathrm{eq}$ = 2300~K (blue) and $|(S/N)_{\Delta \lambda}|_\mathrm{min} \simeq$  {10} at 4~$\mathrm{\mu m}$ for $T_\mathrm{eq}$ = 3000~K (red). 
 Thus, observation of an HRSE in secondary eclipse can detect the components of the mineral atmosphere such as Na, K and SiO, provided its signal to noise ratio is larger than $|(S/N)_{\Delta \lambda}|_\mathrm{min}$.

\begin{figure}[tbp]
   \begin{center}
\includegraphics[width=90mm]{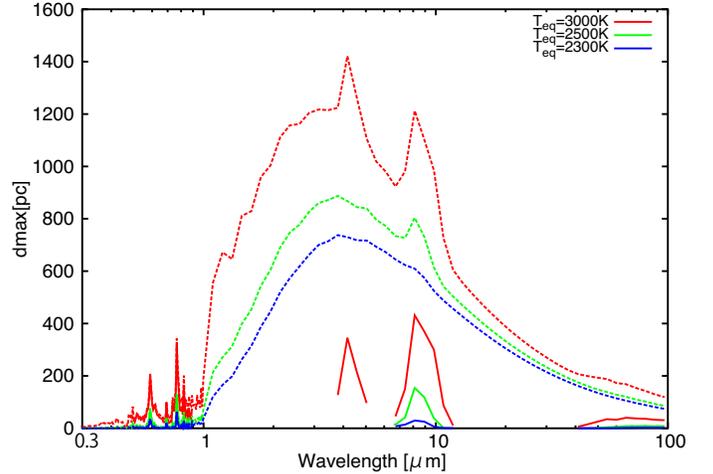}
   \caption{The limiting distance $d_\mathrm{max}$ (see eq.~[\ref{eq:dmax}] for the 
   definition), interior to which an HRSE with the model atmosphere orbiting 
   a Sun-like star is detectable with $(S/N)_{\Delta \lambda}$ by a given instrument,
    is shown as a function of wavelength 
   for three equilibrium temperatures $T_\mathrm{eq}$ = 3000~K (red), 
   2500~K (green), and 2300~K (blue) with the assumption that 
   $R_*$ = 1~$R_{\odot}$, $R_p$ = $0.02 R_{\odot}$, $A=\pi (5 \mathrm{m}/ 2)^2$, 
   $t_\mathrm{obs}$ = 10~hours, and \Pisymbol{psy}{"C2} = 10 in 1-100 $\mathrm{\mu m}$ 
   and 100 in 0.3-1 $\mathrm{\mu m}$. 
   The corresponding atmospheric structure is shown 
   in Fig.~\ref{fig:HSE_TP_toon}.
   The solid lines show the limiting distance $d_\mathrm{max}$ to detect the line spectral features within secondary eclipse depth around 4, 10, and 100 $\mathrm{\mu m}$,
    assuming $(S/N)_{\Delta \lambda}= 3 / \eta$ shown in Figure~\ref{fig:snl}b.
    The dotted lines show the limiting distance $d_\mathrm{max}$ to detect the secondary eclipse depth itself, assuming $(S/N)_{\Delta \lambda}$=3.
   The host star is assumed to be a Sun-like star with radius of 1~$R_\odot$ 
   (i.e., the planetary/stellar radius ratio being 0.02) and 
   emit the blackbody radiation of 6000~K. 
  }
\label{fig:dmax}
\end{center}
\end{figure}

Finally, we estimate the limiting distance $d_\mathrm{max}$
 interior to which an HRSE with the mineral atmosphere is detectable with a given $|(S/N)_{\Delta \lambda}|_\mathrm{min}$,
  for the photon-noise limiting case.
From equation~(\ref{eq:snr}), $d_\mathrm{max}$ is written as 
\begin{eqnarray}
  d_\mathrm{max} \left\{ |(S/N)_{\Delta \lambda}|_\mathrm{min} \right\} &=&\frac{\epsilon_{\lambda, \Delta \lambda}}
  {|(S/N)_{\Delta \lambda}|_\mathrm{min}} \notag
  \\
  && \times \sqrt{\frac{At_\mathrm{obs}}{\mbox{\Pisymbol{psy}{"C2}}}
  \frac{\pi B_{\Delta\lambda}(T_*)}{h}}R_{*}.
  \label{eq:dmax}
\end{eqnarray}

{
Figure~\ref{fig:dmax} shows $d_\mathrm{max}$ as a function of wavelength for
 $|(S/N)_{\Delta \lambda}|_\mathrm{min}$ = $3/\eta_{\lambda,\Delta \lambda}$
  (i.e., the minimum signal-to-noise ratio for the detection of line spectral features within secondary eclipse depth; shown by solid lines)
  or $3$ (i.e., the minimum signal-to-noise ratio for the detection of the secondary eclipse depth itself; shown by dotted lines);
  $R_*$ = 1~$R_{\odot}$, $R_p$ = $0.02 R_{\odot}$, $A=\pi (5 \mathrm{m}/ 2)^2$, $t_\mathrm{obs}$ = 10~hours, and \Pisymbol{psy}{"C2} = 10 in 1-100 $\mathrm{\mu m}$ and 100 in 0.3-1 $\mathrm{\mu m}$. 
   In particular, the solid lines show the limiting distance $d_\mathrm{max}$ to detect the SiO line spectral features around 4, 10, and 100 $\mathrm{\mu m}$.
   Note that $d_\mathrm{max}$ around 4 $\mathrm{\mu m}$ is not shown for $T_\mathrm{eq}$ = 2300~K and 2500~K, 
   because the Na, K and SiO features in those cases are indistinguishable from each other
   when integrating the secondary eclipse depth {with respect to} wavelength with \Pisymbol{psy}{"C2} = 10.
   }

  Table \ref{tbl:dmax} summarizes $d_\mathrm{max}$ for the SiO detection. 
 The  limiting distance $d_\mathrm{max}$ increases with increasing $T_\mathrm{eq}$,
  because {an HRSE with  higher $T_\mathrm{eq}$ shows 
  greater planetary emission and more remarkable line absorption spectral feature} of the secondary eclipse depths.
 Thus, the mineral atmospheres of more distant HRSEs would be also detectable if they have higher $T_\mathrm{eq}$. 
 
 \begin{table}[htbp]
\caption{Limiting distance {for detection of the SiO line features}}
    \label{tbl:dmax}
      \begin{center}
  \begin{tabular}{ l r r r } \hline \hline
    $d_{\mathrm{max}}$ & $T_\mathrm{eq}$ = 2300~K & $T_\mathrm{eq}$ = 2500~K & $T_\mathrm{eq}$ = 3000~K  \\ \hline 
    4$\mathrm{\mu m}$ & - & - & 340pc  \\ 
  10$\mathrm{\mu m}$ & 30pc & 150pc & 430pc  \\
    100$\mathrm{\mu m}$ & 3.5pc & 8.7pc & 40pc \\ \hline
  \end{tabular}  
    \end{center}
%  The limiting distances for the photon noise limit for $T_\mathrm{eq}$ = 2300~K, $T_\mathrm{eq}$ = 2500~K and $T_\mathrm{eq}$ = 3000~K.  These distances are required to detect the SiO line spectra feature around 4, 10, 100 $\mathrm{\mu m}$   with the assumptions that $R_*$ = 1~$R_{\odot}$, $R_p$ = $0.02 R_{\odot}$, $A=\pi (5 \mathrm{m}/ 2)^2$, $t_\mathrm{obs}$ = 10~hours and \Pisymbol{psy}{"C2} = 10.
\end{table}
 
%
% Section 5.3
%

\subsection{{Application to known HRSEs} \label{subset:hrse_exam}}
\begin{figure}[tbp]
   \begin{center}
  \includegraphics[width=80mm]{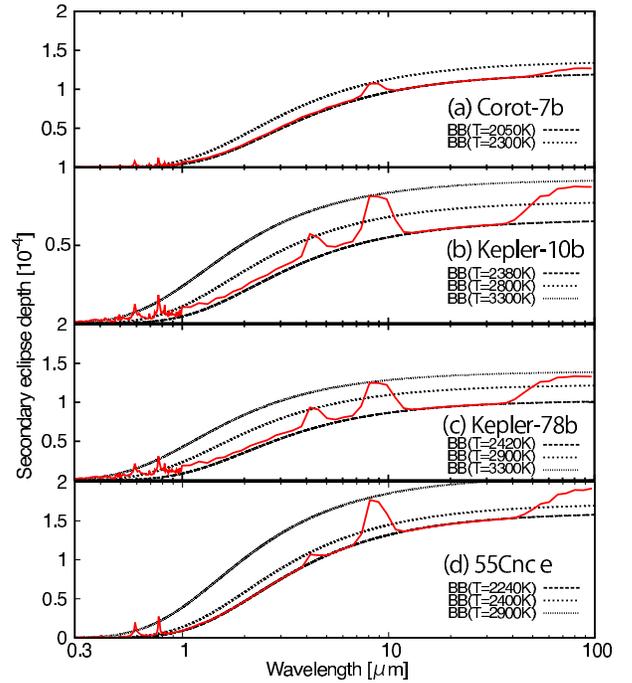}
   \end{center}
   \caption{The secondary eclipse depths theoretically predicted for (a) Corot-7~b, (b) Kepler-10~b, (c) Kepler-78~b and (d) 55 Cnc e, shown as a function of wavelength, assuming that they have the mineral atmosphere.
  The solid lines show the predicted secondary eclipse depth.
  The black dotted lines show the secondary eclipse depths of the blackbody, 
  temperatures of which are indicated by ''BB($T$)''. 
  We have assumed \Pisymbol{psy}{"C2} = 100 in 0.3-1 $\mathrm{\mu m}$ and 10 in 1-100 $\mathrm{\mu m}$
   and the planets' parameters shown in Table~\ref{tbl:hrsemr}.
  }
\label{fig:ex_hrse}
\end{figure}

{We show the simulated secondary-eclipse-depth spectra of four known close-in super-Earths, CoRoT-7b, Kepler-10b, Kepler-78b, and 55~Cnc~e in Fig.~\ref{fig:ex_hrse}. 
Within the 1$\sigma$ errors of their observed masses and radii, it is possible that those super-Earths are naked rocky planets. 
Note that, according to \citet{Dragomir+2014},
 the 1$\sigma$ upper limit on the measured 55~Cnc~e's density just reaches the pure rock regime.
If so, they have mineral atmospheres, because of high $T_\mathrm{eq}$ (see Table~\ref{tbl:hrsemr}). 
In the simulations, we have assumed \Pisymbol{psy}{"C2} = 10 in 1-100 $\mathrm{\mu m}$ and 100 in 0.3-1 $\mathrm{\mu m}$.
The values of the planetary and stellar parameters that we have used are listed in Table~\ref{tbl:hrsemr}. 
}

{
The secondary eclipse depth becomes large, as $R_p$/$R_\ast$ increases and $T_\mathrm{eq}$ increases. 
Thus, the secondary eclipse depth of 55 Cnc~e ($R_p/R_\ast$ = 2.12 $\Rearth$/$R_{\odot}$) is largest, while that of Kepler-10b ($R_p/R_\ast$ = 1.39  $\Rearth$/$R_{\odot}$) is smallest. 
Despite of its larger $R_p/R_\ast$, the secondary eclipse depth of CoRoT-7b is smaller than that of Kepler-78b, because $T_\mathrm{eq}$ of CoRoT-7b is smaller than that of Kepler-78b. 
}

{
Prominent absorption features are found around $\lambda =$ 0.6~$\mathrm{\mu m}$ (Na), 0.77~$\mathrm{\mu m}$ (K), 4~$\mathrm{\mu m}$, 10~$\mathrm{\mu m}$ and 100~$\mathrm{\mu m}$ (SiO). 
In particular, in the cases of Kepler-10b and 55~Cnc~e, those features would be detectable by the near-future space-based observation that we have assumed in subsection~\ref{subsec:dep_r2}, because the distances of these systems are shorter than ($R_*$/$R_{\odot}$)$^{-1}$ ($R_p$/$2 \Rearth$)$^2$ $d_{\mathrm{max}}$ 
from equations~(\ref{eq:dp}) and (\ref{eq:dmax}).
}

\begin{table}[tbp]
\caption{Assumed planetary and stellar parameters}
    \label{tbl:hrsemr}
      \begin{center}
  \begin{tabular}{ l r r r r } \hline \hline
     & Corot-7b$^a$ & Kepler-10b$^b$ & Kepler-78b$^c$ & 55 Cnc e$^d$  \\ \hline 
    $M_p$ & 7.42 $\Mearth$ & 3.33 $\Mearth$ & 1.69 $\Mearth$ & 8.09 $\Mearth$ \\
    $R_p$ & 1.58 $\Rearth$ & 1.47 $\Rearth$ & 1.20 $\Rearth$ & 1.99 $\Rearth$ \\ 
    $D$ & $0.0172$AU & 0.0168 AU & 0.0092 AU & 0.0155 AU \\
    $T_\mathrm{eq}$ & 2550~K & 3090~K & 3000~K & 2700~K\\
    $R_*$  & 0.82~$\Rsun$ & 1.06~$\Rsun$ & 0.74~$\Rsun$ & 0.94~$\Rsun$ \\
    $T_*$  & 5250~K & 5700~K & 5200~K & 5200~K \\ 
    distance & 150pc & 173pc & - & 12.3pc \\ \hline
  \end{tabular}  
  \\
  Notes.
  \footnotemark{\citet{Hatzes+2011,Bruntt+2010,Leger+09}}, \footnotemark{\citet{Dumusque+2014,Fressin+2011}}, \footnotemark{\citet{Howard+2013}}, \footnotemark{\citet{Dragomir+2014,Nelson+2014,Ehrenreich+2012}}
    \end{center}
\end{table}

{
\citet{Demory+2012} have already reported the detection of the secondary eclipse depth of 55 Cnc~e with Spitzer Space Telescope at 4.5 $\mu$m.
The measured brightness temperature of 2360 $\pm$ 300~K is consistent with our result shown in Fig.~\ref{fig:ex_hrse}(d), although the error is large.
Also, the SiO IR-absorption features of an HRSE that is brighter and closer than 55 Cnc e would be detectable with Spitzer in the IRAC 4.5 and 8 $\mu$m channel.
}
  
  \begin{table*}[t]
\caption{{Values of the line contrast}}
    \label{tbl:siosign}
\begin{center}
        \scalebox{1.0}[1.0]{
  \begin{tabular}{ l r r r r r } \hline \hline
    the line contrast $C_{L,(\lambda,\Delta \lambda)}$ & $T_\mathrm{eq}$ = 2300~K & $T_\mathrm{eq}$ = 2500~K & $T_\mathrm{eq}$ = 3000~K & 55 Cnc e & Accuracy of EChO  \\ \hline 
    Na at 0.59~$\mathrm{\mu m}$ & 4.4ppm  & 9.3ppm & 26ppm & 19ppm (\Pisymbol{psy}{"C2} = 100) & -   \\ 
    K at 0.77~$\mathrm{\mu m}$ & 6.4ppm & 14ppm & 36ppm & 26ppm (\Pisymbol{psy}{"C2} = 100) & -  \\ 
    SiO at 4~$\mathrm{\mu m}$ & - & - & 30ppm & 16ppm (\Pisymbol{psy}{"C2} = 50) & 12ppm (\Pisymbol{psy}{"C2} = 50)  \\ 
  SiO at 10~$\mathrm{\mu m}$ & 4.6ppm & 24ppm & 67ppm & 54ppm (\Pisymbol{psy}{"C2} = 30) & 24ppm (\Pisymbol{psy}{"C2} = 30) \\
    SiO at 100~$\mathrm{\mu m}$ & 4.0ppm & 13ppm & 50ppm & 36ppm (\Pisymbol{psy}{"C2} = 30) & - \\ \hline
  \end{tabular}  
  }
  \end{center} 
 % The line contrasts, $C_L = \eta \, \epsilon_{\lambda,  \Delta \lambda}$, are shown for $T_\mathrm{eq}$ = 2300~K, $T_\mathrm{eq}$ = 2500~K, and $T_\mathrm{eq}$ = 3000~K (\Pisymbol{psy}{"C2} = 100 for $\lambda \le$ 1~$\mathrm{\mu m}$ and \Pisymbol{psy}{"C2} = 10 for $\lambda \ge$ 1~$\mathrm{\mu m}$) and  55 Cnc e (\Pisymbol{psy}{"C2} = 100 for $\lambda \le$ 1~$\mathrm{\mu m}$, \Pisymbol{psy}{"C2} = 50 for 1~$\mathrm{\mu m}$ $\le$ $\lambda \le$ 5~$\mathrm{\mu m}$ and \Pisymbol{psy}{"C2} = 30 for $\lambda \ge$ 5~$\mathrm{\mu m}$). The photometric accuracy of EChO in Chemical Census mode of 55 Cnc e \citep{Waldmann+14} is also shown.  These signals include Na (0.59~$\mathrm{\mu m}$), K (0.77~$\mathrm{\mu m}$) and SiO (4~$\mathrm{\mu m}$, 10~$\mathrm{\mu m}$ and 100~$\mathrm{\mu m}$).
\end{table*}
  
\begin{figure}[htbp]
   \begin{center}
  \includegraphics[width=90mm]{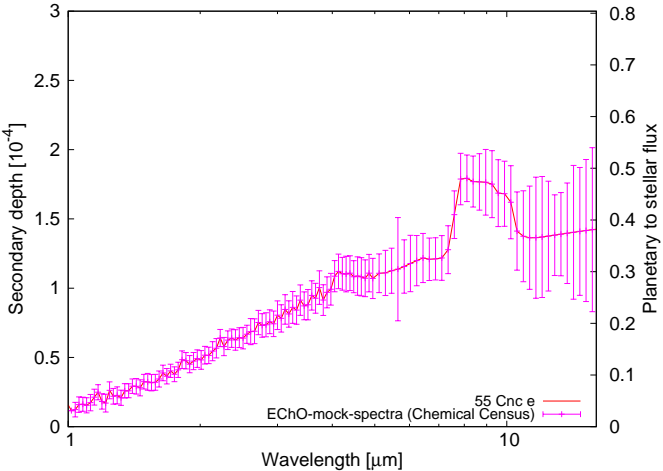}
   \end{center}
   \caption{The secondary eclipse depths theoretically predicted and the mock-EChO
  observational spectrum for 55 Cnc e with the mineral atmosphere.
  The secondary eclipse depth is shown as a function of wavelength.
  The solid line shows the predicted secondary eclipse depth.
  The magenta bars show the mock-EChO observational spectrum \citep{Waldmann+14}.
  We have assumed \Pisymbol{psy}{"C2} = 50 in 1-5 $\mathrm{\mu m}$ and 30 in 5-16 $\mathrm{\mu m}$
   and 55 Cnc e's parameters: $M_p$ = 8.09 $\Mearth$, $R_p$ = 1.99 $\Rearth$, $a$ = $0.0155$AU, $T_\mathrm{eq}$ = 2700~K
    $R_*$=0.94~$\Rsun$ and $T_*$=5200K.     
  }
\label{fig:55cnce_echo}
\end{figure}

Using equations (\ref{eq:snr}) and (\ref{eq:eta}), we define the line contrast to the total flux,
  $C_{L,(\lambda,\Delta \lambda)}$ as
  \begin{equation}
 C_{L,(\lambda,\Delta \lambda)} \equiv \eta_{\lambda,\Delta \lambda} \, \epsilon_{\lambda, \Delta \lambda}.
\end{equation}
   The photometric accuracy should be smaller than $C_{L,(\lambda,\Delta \lambda)}$ in order to detect the line spectra feature of the secondary eclipse depth.
    There are several proposed missions that aim to characterize transiting exoplanets, such as Fast Infrared Exoplanet Spectroscopy Survey Explorer \citep[FINESSE; e.g.][]{2012SPIE.8442E..41D} and the Exoplanet Characterization Observatory \citep[EChO; e.g.][]{Tinetti+12}. 
   Here we compare our results with the accuracy of EChO. Figure \ref{fig:55cnce_echo} shows the theoretical (solid line) and mock-EChO (bars) secondary-eclipse-depth spectra for a super-Earth with a mineral atmosphere, using 55 Cnc e's parameters.
    Then, we create the mock-EChO spectra with \Pisymbol{psy}{"C2} of 50 in 1--5 $\mathrm{\mu m}$ and 30 in 5--16 $\mathrm{\mu m}$, and the photometric accuracy of EChO in Chemical census mode \citep{Waldmann+14} \footnote{\citet{Waldmann+14} showed the simulated EChO observation of 55 Cnc e assuming a CO$_2$ and H$_2$O atmosphere model. We use the noise level assumed in their Figure 9 because the depth of their spectra is similar to ours.}.  Our result demonstrates that EChO has enough accuracy to detect the SiO feature of the mineral atmosphere of 55 Cnc e around 4 $\mathrm{\mu m}$ and 10 $\mathrm{\mu m}$. 
    
   Table \ref{tbl:siosign} shows the line contrasts $C_L$ of Na (0.59~$\mathrm{\mu m}$), K (0.77~$\mathrm{\mu m}$) and SiO (4~$\mathrm{\mu m}$, 10~$\mathrm{\mu m}$ and 100~$\mathrm{\mu m}$) for $T_\mathrm{eq}$ = 2300~K, 2500~K, and 3000~K (\Pisymbol{psy}{"C2} = 100 for $\lambda \le$ 1~$\mathrm{\mu m}$ and \Pisymbol{psy}{"C2} = 10 for $\lambda \ge$ 1~$\mathrm{\mu m}$), and 55 Cnc e's parameters  (\Pisymbol{psy}{"C2} = 100 for $\lambda \le$ 1~$\mathrm{\mu m}$, \Pisymbol{psy}{"C2} = 50 for 1~$\mathrm{\mu m}$ $\le$ $\lambda \le$ 5~$\mathrm{\mu m}$ and \Pisymbol{psy}{"C2} = 30 for $\lambda \ge$ 5~$\mathrm{\mu m}$). The photometric accuracy of EChO in Chemical Census mode of 55 Cnc e \citep{Waldmann+14} is also shown in  Table \ref{tbl:siosign}. 
      {
      Note that {the values of} $C_L$ at 4 $\mathrm{\mu m}$ are not shown for $T_\mathrm{eq}$ = 2300~K and 2500~K, 
   because the Na, K and SiO features in those cases are indistinguishable with \Pisymbol{psy}{"C2} = 10.
   }
   We obtain that $C_{L,(\lambda,\Delta \lambda)}$ $\gtrsim$ 10 \rm{ppm} for Na (0.59~$\mathrm{\mu m}$), K (0.77~$\mathrm{\mu m}$) and SiO (4~$\mathrm{\mu m}$, 10~$\mathrm{\mu m}$ and 100~$\mathrm{\mu m}$); for example, $C_{L,(\lambda,\Delta \lambda)}$ $\sim$ {14} ppm at 0.77~$\mathrm{\mu m}$ and {24} ppm at 10~$\mathrm{\mu m}$ for $T_\mathrm{eq}$ = 2500~K and $C_{L,(\lambda,\Delta \lambda)}$ $\sim$ {26} ppm at 0.59~$\mathrm{\mu m}$, {34} ppm at 0.77~$\mathrm{\mu m}$, and {67} ppm at 10~$\mathrm{\mu m}$ for $T_\mathrm{eq}$ = 3000~K.
The required photometric accuracy to detect the SiO features at 4 and 10~$\mathrm{\mu m}$ is achievable with the Chemical Census mode of EChO for 55 Cnc e.

%
% Section 5.4
%
\subsection{How many HRSEs are detectable via near-future observation? \label{sec:diss_hrse}}

The {\it Kepler} space telescope discovered {4250} planet candidates {(http://exoplanets.org, as of 30 Nov. 2014)}. 
The Kepler Objects of Interest (KOIs) include super-Earth-size objects with short periods (i.e., HRSEs), in particular, {751} objects with $R_p \leq 2 \Rearth$ orbiting stars with $T_\mathrm{eff}$ = 5300-6000~K. Among them, there are {60} KOIs with $T_\mathrm{eq} >$ 2300~K. If they are naked rocky planets, their mineral atmospheres can be identified by secondary-eclipse observations with sufficient signal-to-noise ratios. We estimate the expected number of such HRSEs detectable by near future observations.

Assuming a uniform distribution of G-type stars, 
the number of the detectable HRSEs, $\overline{N_\mathrm{HRSE}}$, is given by
\begin{equation}
	\overline{N_\mathrm{HRSE}} \approx
		r N_\mathrm{G} \frac{4\pi}{3} \left[ d_\mathrm{max} \left\{ (S/N)_{\Delta \lambda} \right\} \right]^3,
 \label{eq:nhrse}
\end{equation}
{
where $r$ is the occurrence rate of HRSEs with $T_\mathrm{eq} \geq 2300$~K hosted by G-type stars }
 and $N_\mathrm{G}$ is the number density of G-type stars. 
  The HRSEs with $T_\mathrm{eq} \geq 2300$~K, which should exhibit the prominent emission features (Fig.~\ref{fig:mockdep}),
 correspond to planets with period $P \lesssim 2$~days around G-type stars.
  \citet{Fressin+2013} estimated, using KOI host stars, the occurrence rate of planets with $1.25 R_{\oplus} < R_p < 2R_{\oplus}$ hosted by G-type stars to be close to $0.35$. They also estimated the occurrence rate of planets with $1.25 R_{\oplus} < R_p < 2R_{\oplus}$ and $0.8 < P < 2$ days to be $1.7 (\pm 0.3) \times 10^{-3}$.  Using their results, the occurrence rate of the planets with $1.25 R_{\oplus} < R_p < 2R_{\oplus}$ and $0.8 < P < 2$ days hosted by G-type stars is estimated to be $r \sim 6 \times 10^{-4}$. Here we use $r$ = $6 \times 10^{-4}$ as a fiducial value of the occurrence rate.

In section~\ref{subsec:dep_r2}, we estimated ${d_\mathrm{max}}\left\{ (S/N)_{\Delta\lambda} = 3/\eta_{\lambda,\Delta \lambda}  \right\}$ to be {430~pc} ($T_\mathrm{eq}$ = 3000~K), {150~pc} ($T_\mathrm{eq}$ = 2500~K) and {30~pc} ($T_\mathrm{eq}$ = 2300~K) for 10~$\mathrm{\mu m}$ observation (see Fig.~\ref{fig:dmax} and Table~\ref{tbl:dmax}).  These HRSEs with $T_\mathrm{eq}=2300$~K, $2500$~K and $3000$~K have the orbital period $P \sim 0.93$, 1.6 and 2.0~days, respectively. Hence we take the simple mean $\left[ ( 430^3 + 150^3 + 30^3 )/3 \right]^{1/3} \rm{pc} \sim 300\,{\rm pc}$ as a fiducial value of the limiting distance for HRSEs with $0.8 < P < 2$ days. Inputting $r = 6 \times 10^{-4}$ \% and $N_\mathrm{G}$ $\sim$ $6$$ \times 10^{-3} \mathrm{pc}^{-3}$ \citep[see a recent review by][]{Traub+11} in equation~(\ref{eq:nhrse}), we find that within {300} pc there are approximately {400} HRSEs whose 10~$\mathrm{\mu m}$ SiO features are detectable .  Taking the transiting probability of $\sim$ 0.2 ($\sim\Rsun/D$ for $P =$ 1.3 days) into account, one will be able to detect the 10~$\mathrm{\mu m}$ SiO features from $\sim$ {80} transiting HRSEs by a 5~m space telescope equipped with a photon-noise limited instrument.

%%
%% Section 6
%%
\section{MODEL VALIDATION AND FUTURE STUDY} \label{sec:discussion}
\subsection{Scattering} \label{Scattering}
In this study, we have ignored the effect of scattering on the thermal structure of the mineral atmosphere. There are two possible sources for scattering. One is Rayleigh scattering by the major gases. 
The cross section for Rayleigh-scattering by gas species A, $\sigma_{s,\nu}^A$, 
ignoring the polarizability anisotropy,
 are approximately given as \citep[e.g.,][]{CRC92}
\begin{equation}
	\sigma_{s,\nu}^A=\frac{8\pi}{3c^4}(2\pi\nu)^4\alpha_A^2,
\label{eq:rs}
\end{equation}
where $\alpha_A$ is the static average electric dipole polarizability of gas species A. 
According to \cite{CRC92}, $\alpha_\mathrm{A}$ for ground-state atoms and diatomic molecules in the unit of $10^{-24}$~cm$^3$ are 
24.11 for Na, 
43.4 for K, 
0.802 for O, 
and 1.5689 for O$_2$. While $\alpha_\mathrm{A}$ for SiO is unavailable in the literature, to our knowledge, it should not differ by several orders of magnitude from those of the other gases. 
Comparing the cross sections for Rayleigh scattering by those gases with the total absorption cross section calculated in section~\ref{sec:opacity}, we have confirmed that the formers are smaller by several orders of magnitude than the latter in most of the wavelength region of interest, except for the narrow ranges between 0.1 and 0.15~$\mathrm{\mu m}$ and between 0.3 and 0.45~$\mathrm{\mu m}$.
We have integrated the thermal structure incorporating the effect of the Rayleigh scattering by Na, {which results in only} {a few percent} difference in temperature. 
Also, the Rayleigh and Mie scattering by cloud particles can affect the thermal structure of the mineral atmosphere, if present. 
However, since {the atmospheric temperature is always higher than the vapor pressure of rock, as} shown in Fig.~\ref{fig:HSE_TP_toon}, no condensation cloud is formed in the atmosphere.
 
\subsection{Non-LTE effect \label{subsec:nonlte}}
The approximation of local thermal equilibrium (LTE) is invalid at low pressures.
In this study, we have considered a pressure range of $> 1 \times 10^{-8}$~bar. 
For LTE to be attained, thermalization due to atomic/molecular collisions must occur faster than radiative cooling. 
The collisional timescale, $\tau_\mathrm{coll}$, is $\sim$  $kT/(P v \sigma_\mathrm{coll})$, where $\sigma_{coll}$ and $v$ are the collisional cross section and thermal velocity of molecules. 
For SiO, since $\sigma_\mathrm{coll}$ $\approx$ $3 \times 10^{-15}$~cm$^2$, 
$T$ $\approx$ 3000~K, 
$v$ $\approx$ $8 \times 10^4$~cm/s and
the typical radiative-cooling timescale (i.e., the typical lifetime of excited states) $\tau_\mathrm{cool}$ $\approx$ $1\times10^{-2}$~s \citep{Barton+13}, 
LTE is valid for $P \gtrsim 2 \times 10^{-7}$~bar.
Since the emission from the low pressure region of $P \lesssim 2 \times 10^{-7}$~bar makes little contribution to 
the secondary eclipse depth,
 the LTE approximation is valid for the SiO-dominated atmosphere of special interest in this study. 

\subsection{Photodissociation of SiO} \label{subsec:photod}
{Our equilibrium calculations show that} the major gas species in the {mineral} atmosphere are Na, K, Fe, O, O$_2$, Si and SiO (section~\ref{sec:structure}).
{Although having been ignored in this study, photodissociation of SiO would occur in such an atmosphere by irradiation of stellar UV \citep[e.g.,][]{Vorypaev1981}.}

{
\citet{Jolicard+97} provided the photodissociation cross section of SiO, $\sigma_\mathrm{dis}$, in the wavelength range between 0.153~$\mathrm{\mu m}$ and 0.157~$\mathrm{\mu m}$, where the cross section decreases monotonically with increasing wavelength from $\sim$10$^{-17}$ cm$^2$ to 10$^{-19}$ cm$^2$. 
{The energy of UV rays of $\lambda >$ 0.157~$\mathrm{\mu m}$ is not high enough to dissociate SiO significantly; instead, the electron-transition of SiO is dominant, as discussed in section~\ref{sec:opacity}.}
%Since photodissociation cross section does not increase with wavelength in general, $\sigma_\mathrm{dis}$ should be $\gtrsim 10^{-17}$~cm$^2$ for $\lambda <$ 0.153~$\mathrm{\mu m}$ and $\lesssim 10^{-19}$~cm$^2$ for $\lambda >$ 0.157~$\mathrm{\mu m}$. 
%In the mineral atmosphere, for $\lambda =$ 0.157-0.3~$\mathrm{\mu m}$, the electron-transition cross sections of SiO are larger than 10$^{-19}$ cm$^2$ and thus larger than $\sigma_\mathrm{dis}$,as shown in Fig~\ref{fig:cross}.
Also, for $\lambda =$ 0.152-0.163~$\mathrm{\mu m}$, {the electron-transition cross sectio}n of Si is larger than 10$^{-17}$ cm$^2$ and thus larger than $\sigma_\mathrm{dis}$. 
Therefore, it is the UV ray of $\lambda \leq$ 0.152 $\mathrm{\mu m}$ that dissociates SiO efficiently in the mineral atmosphere.
{Although no other published data for $\sigma_\mathrm{dis}$ is available, $\sigma_\mathrm{dis}$ is unlikely to decrease with decreasing wavelength.
Below, we make a rough estimate on the impact of SiO photodissociation on the atmospheric structure and the emission features.}
}

{
Using $\sigma_\mathrm{dis}$ $>$ 1 $\times$ 10$^{-17}$ cm$^2$ with $x_\mathrm{SiO}$ = $1 \times 10^{-3}$, $\overline{\mu}$ = $20$ and $g$ = 2500 cm/s$^{-2}$, 
we calculate the SiO-photodissociation optical depth, which is defined by 
\begin{equation}
	\tau_\mathrm{dis} = \frac{x_\mathrm{SiO} \sigma_\mathrm{dis}}
					   {\overline{\mu} m_\mathrm{H} g}P,
\end{equation}
where $m_\mathrm{H}$ is the proton mass. 
Then, we find that $\tau_\mathrm{dis} = 1$ at $P$ $< 8 \times 10^{-6}$ bar. 
This means that the atmospheric structure for $P \lesssim 8 \times 10^{-6}$~bar is affected significantly by photodissociation of SiO. 
Since the emission features shown in Fig.~\ref{fig:HSE_TP_toon} arise from the region of $P \gtrsim 3 \times 10^{-5}$~bar (see Fig.~\ref{fig:HSE_dep}), the emission spectra are unlikely to be affected by the photodissociation. 
More detailed examination requires photo-chemical calculations done consistently with radiative transfer calculations, which must be a future study.
}

\subsection{Tidal heating}\label{subsec:tidal}
{
Our atmospheric models do not include the effect of tidal heating, discribed in section~\ref{sec:structure} ($F_0$ = 10~W/m$^2$ being assumed). 
According to \citet{Henning+2009}, tidal heating dominates stellar irradiation under somewhat extreme circumstances such that the planetary eccentricity, $e$, is larger than 0.1 and the orbital period is less than 2 days.
Also, \citet{Miguel+11} showed that addition of tidal heating hardly changes the surface temperature of HRSEs for $e$ = 0.01 and 0.001 and the quality factor of the planet, $Q$, of 200 and 20000. 
Indeed, although not shown, 
we have confirmed that there is no significant change in the atmospheric structure for $T_\mathrm{eq}$ = 3000~K, by calculating the temperature profiles with $F_0=10^3$ - $10^7$ W/m$^2$, being equivalent to tidal heat flux for $e$ = 0.01 and 0.001 and $Q$ = 200 and 20000. }

\subsection{Temperature jump at the bottom \label{sec:temp_jump}}
Allowing the temperature jump would affect the composition of the atmosphere.
 If we included the thermal conduction, we would have no temperature jump at the bottom in the atmosphere, and would have higher $T_\mathrm{g}$ and a thicker atmosphere than our model in this study, which is more detectable.
 The HRSE atmosphere is thin enough for radiation to heat the atmosphere more effectively than thermal conduction.
 Thus, although the thermal conduction would change $T_\mathrm{g}$ and thus the composition of the atmosphere, it would be ineffective in changing the emission spectra of HRSEs. 

\subsection{Impact of potassium on thermal structure and detectability \label{sec:diss_k}}
As shown in section~\ref{sec:chemistry}, our chemical-equilibrium calculations yield much more abundant potassium than \citet{Schaefer+09} did. 
Although not shown, we have confirmed that such difference in K abundance has only a small impact on the thermal structure of the atmosphere: 
Our calculations have demonstrated that the maximum differences in temperature are $\sim$2.86\%, $\sim$2.27\% and $\sim$1.89\% for $T_\mathrm{eq}$ = 3000~K, 2500~K and 2300~K, respectively. 
Thus, the secondary eclipse depths in the IR region are hardly affected. 

{
A direct impact on the secondary eclipse depth is found at $\lambda$ = 0.77~$\mathrm{\mu m}$. Detecting potassium may be useful for constraining the magma composition, because the potassium abundance is different by one order of magnitude from MORB to the upper crust, as shown in Figure~\ref{fig:COMP_X}. 
}

\subsection{Other observation methods to detect mineral atmospheres\label{sec:obsm}}
{
Other observation methods for atmospheres of close-in exoplanets include  transmission spectroscopy and  planetary radial velocity (PRV) measurement, in addition to secondary eclipse observation.
It would be, however, difficult to detect the HRSEs' atmospheres via transmission spectroscopy of the planetary disk
 because of their small scale height.
The part of the HRSE's atmosphere observed by transmission spectroscopy
 is the rim where horizontal heat and composition transport between the day side and night side would be important.
The examination requires 3-D calculations of atmospheric circulation done consistently with radiative transfer calculations, which must be a future study.
}

{
Recently, several molecules in the dayside spectra of hot Jupiters have been successfully detected using the high dispersion spectrograph \citep{brogi+2012,Rodler+2012,deKoK+2013,Rodler+2013,Birkby+2013,Lockwood+2014}. This method utilizes the cross-correlation analysis of the spectra and the model template and directly detects PRV. Here, we estimate the feasibility of the PRV detection of the mineral atmosphere of HRSEs. For the most optimistic case, the signal-to-noise ratio of the PRV is determined by the photon noise,
\begin{eqnarray}
(S/N)_\mathrm{PRV} = \sqrt{N_\mathrm{line}\zeta} (S/N)_{\Delta \lambda} C_{L,(\lambda,\Delta \lambda)}, 
\end{eqnarray}
where $N_\mathrm{line}$ is the effective number of molecular lines available for the detection and $\zeta$ is the total efficiency of the high-dispersion instrument \citep[see ][]{Kawahara+2014}. Although the determination of the precise value of $N_\mathrm{line}$ requires full simulations of the spectra including the terrestrial transmission and the night airglow, we here perform 1st order-of-magnitude estimate of the detectability of the Na and K lines here, assuming a future 30 m-class ground-based telescope. 
}

{
The number of Na and K lines within the secondary eclipse depth spectrum of an HRSE for $T_\mathrm{eq}$ $\geq$ 2300K, $N_\mathrm{line}$, is $\sim 100$,  the line contrast, $C_{L,(\lambda,\Delta \lambda)}$, is $\sim 0.5$ and  the mean value of those depths, $\epsilon_\lambda$ is $\sim 10^{-4}$, in the IR wavelength region~(1-10$\mu$m), as shown in Fig.~\ref{fig:HSE_dep}.   
  The resolving power of the instrument, \mbox{\Pisymbol{psy}{"C2}}, is required to be large enough to resolve those Na and K lines.  To know this value of \mbox{\Pisymbol{psy}{"C2}}, we have reintegrated the secondary eclipse depth for $T_\mathrm{eq}$ $=$ 2500K around only several Na and K line peaks in the IR wavelength region~(3.70-3.75$\mu$m) by line-by-line calculation. We have confirmed that $\mbox{\Pisymbol{psy}{"C2}}=10^5$ is enough to resolve the Na and K lines. Thus, we use $\mbox{\Pisymbol{psy}{"C2}}=10^5$ below as a fiducial value of \mbox{\Pisymbol{psy}{"C2}}. From equation (\ref{eq:nhrse}), the distance of the system within which at least one transiting HRSE exists is $\sim 70$pc. Assuming a 30 m telescope with $\zeta=0.1$, we obtain the expected S/N for detecting the Na and K lines,
\begin{eqnarray}
&&(S/N)_\mathrm{PRV} \sim 8 
\left(
\frac{\epsilon_\lambda}{10^{-4}}
\frac{70\mathrm{pc}}{d}
\frac{R_*}{1R_{\odot}}
\frac{C_{L,(\lambda,\Delta \lambda)}}{0.5}
\right)
 \notag \\
&&\times 
\left(
\frac{N_\mathrm{line}}{100}
\frac{\zeta}{0.1}
\frac{A}{\pi15^2\mathrm{m}^2}
\frac{t_\mathrm{obs}}{10\mathrm{hr}}
\frac{10^5}{\mbox{\Pisymbol{psy}{"C2}}}
\frac{\pi B(T_\ast)/h}{10^{26}\mathrm{m^{-2}s^{-1}}}
\right)^{\frac{1}{2}},
\end{eqnarray}
where $10^{26} \mathrm{m^{-2}s^{-1}}$ corresponds to the mean value of $\pi B(T_\ast)/h$ in the wavelength region between 1 and 10$\mu$m, assuming $T_\ast=6000$~K. 
This order-of-magnitude estimate suggests that it would be possible to detect Na and K in the HRSEs' atmospheres via PRV measurement in the very large telescope era.
}

\section{SUMMARY AND CONCLUSIONS}
\label{sec:concl}
In this study, we have examined the detectability of hot rocky super-Earths that have atmospheres in the gas/melt equilibrium with the underlying magma oceans via secondary-eclipse observations. 
First, we have done gas-melt equilibrium calculations and showed that the major atmospheric species are Na, K, {Fe,} SiO, O, and O$_2$ (\S~\ref{sec:chemistry}). We have called this atmosphere the mineral atmosphere in this study.

Next, for those species, we have compiled the absorption line data available in literature, and calculated their absorption opacities in the wavelength region  of 0.1--100~$\mathrm{\mu m}$ (\S~\ref{sec:opacity}). We have found that the SiO absorption dominates in the UV and IR wavelength regions, while strong absorption by Na and K is found in the visible region. In this atmospheric gas, the opacity is larger in the UV and visible regions than in the IR region. 

Third, using those opacities, we have integrated the thermal structure of the atmosphere (\S~\ref{sec:structure}). 
The most important finding is that thermal inversion occurs in the atmosphere with relatively high equilibrium temperature (i.e., $\mu_\ast^{1/4} T_\mathrm{eq} \geq 1900$~K), because the UV and visible absorption by SiO dominates the IR absorption by SiO and others. 
The presence of thermal inversion has a crucial impact on the detectability via secondary eclipse observation.

Finally, by calculating the ratio of the planetary and stellar emission during secondary eclipse, we have quantified the detectability  (\S~\ref{sec:detectability}). 
We have found prominent emission features due to SiO at 4, 10, and 100~$\mathrm{\mu m}$. 
Comparing our predicted spectrum with a mock spectrum we made, 
we have demonstrated that the mineral atmosphere of a hot rocky super-Earth of radius 2~$R_\oplus$ with $T_\mathrm{eq} \gtrsim 2300$~K orbiting a solar analog would be detectable by a 5-m space telescope.

The Transiting Exoplanet Survey Satellite (TESS) is a planned space telescope for NASA's Explorer program and is dedicated to an all-sky transit survey of exoplanets \citep{Ricker+10}. 
It is predicted that TESS will discover more than 100 transiting Earth-sized exoplanets with short orbital periods. 
After discovery of nearby HRSEs, follow-up observations by satellites with large telescopes such as JWST, {Spitzer,} FINESSE and EChO will open new exotic worlds of the mineral atmospheres.

\acknowledgements
We thank E. J. Barton and J. Tennyson for sharing their knowledge of SiO opacity with us. We also thank Y. Abe, K. Kurosaki and B. Fegley, Jr. for fruitful discussion. We really appreciate the anonymous referee's careful reading and constructive comments, which has helped improved this paper greatly. This study is supported by the Challenging Research Award from the Tokyo Institute of Technology, Grants-in-Aid for Scientific Research (No. 25400224 and 25800106) from the Japan Society for the Promotion of Science (JSPS), and the Astrobiology Project of the CNSI, NINS (AB261006).
\bibliography{ref}

\appendix
\section{Appendix A: Calculation method of the mean opacities} \label{sec:cal_mo}
 In this study, we use mean opacities for simulating the atmospheric structure and emission spectra, 
 instead of line-by-line calculations. 
 While we integrate the line opacities of O, O$_2$, Na, K and Si with wavelength resolution high enough to reproduce line profiles, 
 we use low resolution grids to calculate the mean opacity of SiO, because there are too many absorption lines 
 to reproduce the line profile of SiO in a practical CPU time.

\begin{figure}[h]
   \begin{center}
\includegraphics[width=100mm]{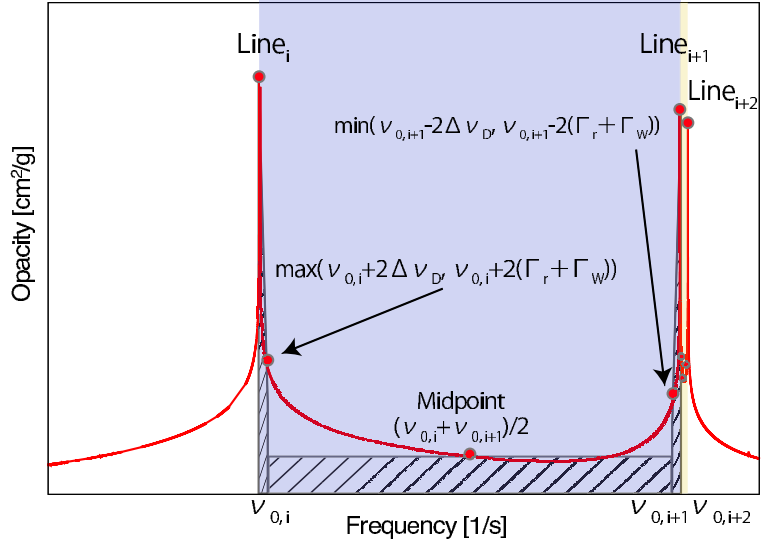}
   \caption{Schematic line profile of SiO (red line) 
   with illustration for selecting grids (red points) to calculate the SiO mean opacity. 
%   The SiO line opacity is shown as a function of wavelength
%   The red lines show the SiO line opacities with enough wavelength resolution to reproduce line profile.
%   The red points shows the SiO line opacities at the line center and its neighborhoods. 
   In the yellow-colored (blue-colored) region, the interval between two adjacent lines is smaller (larger) than $2\Delta \nu_D$ and (or) $2(\Gamma_r+\Gamma_{W})$.
%   In the blue-colored region, the interval between two adjacent lines is larger than $2\Delta \nu_D$ or $2(\Gamma_r+\Gamma_{W})$.
   For example, the opacity integrated between $\nu_{0,i}$ and $\nu_{0,i+1}$ is approximated to be the sum of the areas of three shaded boxes.
  }
\label{fig:opamean}
\end{center}
\end{figure}

 Figure~\ref{fig:opamean} presents how to calculate the mean opacity of SiO when an interval between 
 adjacent lines is less than 1~$\mu$m. 
 When an interval between adjacent lines is smaller than $2\Delta \nu_D$ and $2(\Gamma_r+\Gamma_{W})$ (see the yellow-colored region in Figure~\ref{fig:opamean}),
 we divide the interval equally into quarters. 
 On the other hand, when an interval is larger than $2\Delta \nu_D$ or $2(\Gamma_r+\Gamma_{W})$ (see the blue-colored region in Figure \ref{fig:opamean}), 
 we select a total of five grids: 
 $\nu_{0,i}$, $\nu_{0,i+1}$, the larger of $\nu_{0,i}+2\Delta \nu_D$ and $\nu_{0,i}+2(\Gamma_r+\Gamma_{W})$ ($\equiv \nu_\ell$), 
 the smaller of $\nu_{0,i+1}-2\Delta \nu_D$ and $\nu_{0,i+1}-2(\Gamma_r+\Gamma_{W})$ ($\equiv \nu_r$),
  and a midpoint between $\nu_{0,i}$ and $\nu_{0,i+1}$ ($\equiv \nu_m$), where $\nu_{0,i}$ is the central
frequency of $i$th spectral line.  
  When we integrate the line opacities around the line center, we use a trapezoidal integration method.
  On the other hand, when integrating line opacities between $\nu_\ell$ ($\nu_r$) and $\nu_m$, 
  we approximate the integration as the sum of the areas of the shaded boxes. 
  
  When an interval is larger than 1~$\mu$m, 
  we use the line opacities at $\nu_{0,i}$, $\nu_m$ and 50 grids between those points;  
  those grids are selected so that the size of each interval logarithmically increases with frequency. 
  The integration is done in the same way above. 

 When simulating the atmospheric thermal structure, we adopt the harmonic mean opacities. 
 This is because the left side of equation~(\ref{eq:toon2}) contains the Planck's function $B_\nu (T)$ and 
 its right side contains the opacity in the denominator of $dF^{\pm}_\nu / d\tau_\nu$.
 On the other hand, when calculating the planetary emission spectra, 
 we use the arithmetic mean opacities. 
 This is because equation~(\ref{eq:fp}) includes the Planck's function multiplied by the exponential of the negative opacity and the opacity (i.e., $B_\lambda e^{- \tau_\lambda} d\tau_\lambda$), which can be expressed as the sum of $B_\lambda \kappa_\lambda^m$ ($m = 1, 2, 3, \cdots$). 

%%
%% FIGURES
%%

\end{document}

%% file: rule.tex
 \definecolor{purple}{cmyk}{0.45,0.86,0,0}
 
 \definecolor{brown}{cmyk}{0,0.81,1,0.60}
 
 \newcommand{ \Rsun }{R_\odot} 						%太陽半径
 \newcommand{ \Mearth }{M_{\oplus}}
 \newcommand{ \Rearth }{R_{\oplus}}